\documentclass{pasj00}
\draft

\def\lsim{\mathrel{\mathpalette\Oversim<}}
\def\gsim{\mathrel{\mathpalette\Oversim>}}
\def\Oversim#1#2{\lower0.5ex\vbox{\baselineskip0pt\lineskip0pt%
            \lineskiplimit0pt\ialign{%
          $\mathsurround0pt #1\hfil##\hfil$\crcr#2\crcr\sim\crcr}}}

\begin{document}
\SetRunningHead{H. Mizusawa, K. Omukai, and R. Nishi}
{Primordial Emission in Pop III Galaxies}
\Received{2004/1/1}%{yyyy/mm/dd}
\Accepted{2004/1/1}%{yyyy/mm/dd}

\title{Primordial Molecular Emission 
in Population III Galaxies}

\author{Hiromi \textsc{Mizusawa}}        
\affil{Department of Physics, Niigata University, Ikarashi, Niigata,
 Niigata 950-2181}
\email{mizusawa@astro.sc.niigata-u.ac.jp}
\author{Kazuyuki \textsc{Omukai}}
\affil{National Astronomical Observatory, 
Osawa, Mitaka, Tokyo 181-8588}
\email{omukai@th.nao.ac.jp}
\author{Ryoichi \textsc{Nishi}}     
\affil{Department of Physics, Niigata University, Ikarashi, Niigata,
 Niigata 950-2181}


\maketitle

\begin{abstract}
We study formation of molecules in primordial prestellar 
clumps and evaluate the line luminosities to assess detectability 
by next-generation facilities. 
If the initial H$_2$ fraction is sufficiently high, 
HD becomes an important coolant in the clumps. 
The luminosity from such HD cooling clumps is lower than that from H$_{2}$ 
cooling ones because of lower temperature ($<100$K). 
As for Li reactions, we include the three-body LiH formation 
approximately. 
The Li molecular fraction remains very low ($<10^{-3}$) throughout
the evolution owing to the high dissociative reaction rate of 
${\rm LiH +H \rightarrow Li + H_{2}}$.
LiH does not become an important coolant in any density range. 
The luminous emission lines from the prestellar cores include 
H$_2$ rovibrational lines: 1-0 Q(1), 1-0 O(3), 1-0 O(5), 
and pure rotational lines: 0-0 S(3), 0-0 S(4), 0-0 S(5). 
The next-generation facilities SPICA and JWST are able to 
detect H$_2$ emission in a large pre-galactic cloud that 
forms metal-free stars at a high rate of $\sim 10^3 \MO {\rm yr^{-1}}$
at redshift $z<10$. 
We also derive an analytical expression for the luminosity 
that reproduces the numerical results. 
\end{abstract}

\section{Introduction}
Population III (Pop III) stars, 
which are formed from primordial pristine gas, 
have been postulated as the origin of metals of metal-deficient 
stars in the Galactic halo (Schwarzschild, Spitzer 1953).
Recent detection of very metal-poor stars renewed an interest in 
Pop III stars (e.g. Christlieb et al. 2002).
Pop III stars may also account for the presence of metals 
($\simeq 3.5 \times 10^{-4}Z_{\odot}$) in the high-z ($z\sim 5$) 
L${\rm \alpha}$ absorption systems observed in quasar spectra 
(e.g., Songaila 2001).
Additionally, as a source of ultraviolet photons that 
caused the reionization of the intergalactic medium at $z\simeq 17\pm5$
suggested by WMAP data (Bennett et al. 2003), 
Pop III stars are attracting the attention of many authors
(e.g., Ciardi et al. 2003). 
These observations clearly suggest the importance of roles played by 
Pop III stars in the early evolution of galaxies.

Partly stimulated by the observations, theoretical studies on 
the formation of primordial objects in the early universe have made 
remarkable progress in recent years.
Simultaneously, several next-generation large observational facilities 
are planned, one of whose goals is to detect star formation 
in the first galaxies.
In this context, a variety of observational verifications 
for scenarios of the first object formation 
in the early universe has been proposed.

Detectability of primordial pre-galactic clouds has been 
discussed by some authors.
For example, Haiman, Spaans, and Quataert (2000) and Fardal et al.(2001) estimated 
the L${\rm \alpha}$ luminosity from collapsing pre-galactic clouds 
and discussed their detectability by the James Webb Space Telescope (JWST) 
\footnote{http://www.jwst.nasa.gov/}.
Omukai and Kitayama (2003) demonstrated that 
${\rm H_2}$ lines, in addition to L${\rm \alpha}$ emission, 
were emitted in collapsing pre-galactic clouds and discussed 
detectability of ${\rm H_2}$ rotational lines 
by the Single Aperture Far-Infrared observatory (SAFIR) 
\footnote{ http://safir.jpl.nasa.gov/whatIs/index.asp}. 
${\rm H_2}$ line emission from the supershells produced in 
IGM by supernova (SN) blastwaves 
of Pop III stars can be detected by JWST (Ciardi, Ferrara 2001). 
Observational features of clusters of Pop III stars 
that have already formed have been also studied by many authors 
(e.g., Bromm et al. 2001; Schaerer 2002; Venkatesan et al. 2003).
Near-infrared background anisotropies produced by these clusters 
may have been detected by the Two Micron All Sky Survey (2MASS) 
(e.g., Salvaterra, Ferrara 2003; Kashlinsky et al. 2004).

${\rm H_2}$-line luminosity from individual Pop III star-forming 
regions has also been evaluated (Kamaya, Silk 2002; Ripamonti et al. 2002; 
Mizusawa et al. 2004, hereafter paper I).
At first, the ${\rm H_2}$-line emission from the primordial cloud 
including a number of first star-forming regions was expected to be observable 
by next generation facilities (Kamaya, Silk 2002; Ripamonti et al. 2002).
Later, a flaw was found in the luminosity distance used in their estimations. 
With the correct luminosity distance, 
detectability of the ${\rm H_2}$ emission at the typical 
epoch for first star formation ($z\sim 20$) is extremely low  
even with the Space Infrared Telescope for Cosmology and Astrophysics 
(SPICA) \footnote{http://www.ir.isas.ac.jp/SPICA/},  
which is one of the most sensitive facilities planned in this wavelength
band (paper I).

Nonetheless, even in lower redshift (say, $z\sim 5$) primordial gas 
can be present in objects that are formed with delay
(Scannapieco et al. 2003).
Detection of primordial molecular emission from 
the first episode of star formation in such objects 
should be easier than in genuine first objects in the universe.
In the case of star formation in the first objects, whose mass is 
small ($\sim 10^{6}M_{\odot}$), 
dense star-forming cores are formed with little fragmentation (Bromm et al. 1999,2002; Abel et al. 2000,2002).
On the other hand, in large pre-galactic clouds, 
post-shock cooled sheets, which are expected to form after virialization, 
fragment into filamentary clouds (Yamada, Nishi 1998; Yamada 2002). 
After further contraction and fragmentation of these filaments, 
star-forming cores are produced. 
In such clouds, H$_2$ and HD form abundantly owing to ionization 
accompanying the strong shock (Uehara, Inutsuka 2000; Nakamura, Umemura 2002) 
and strong molecular emission is expected.
Another trace molecule in primordial gas, LiH, can be an important coolant
if all the Li become LiH in spite of small Li abundance (Lepp, Shull 1984).
Luminosities of these trace primordial molecular (i.e., HD and LiH) 
emission were calculated, and their detectability by ALMA was discussed 
by Kamaya and Silk (2003). 
In their estimation, however, they again used the incorrect luminosity 
distances as in Kamaya and Silk (2002). 
Moreover, their assumption that all the deuterium and lithium are in 
molecular form, in particular for Li, is not justified, as described 
later.

In this paper, we investigate the formation of HD and LiH 
as well as H$_2$ not also in collapsing star-forming cores 
but also in filamentary clouds, and estimate their line luminosities.
Here, to assess the importance of HD and LiH lines as coolants, 
we include the deuterium and lithium chemical reactions 
in addition to the processes considered in paper I and 
calculate these molecular fractions and luminosities.
This treatment improves that of Kamaya and Silk (2003), 
who adopted a rather crude assumption for HD and LiH fraction 
without solving chemical reactions.
In particular, we include the Li three-body reaction in an approximate way,
by which the complete conversion of Li into LiH has been envisaged
(Stancil et al. 1996).
Additionally, we consider the formation reaction of LiH
\begin{equation}
{\rm Li}+{\rm H_2}\rightarrow {\rm LiH}+{\rm H},
\end{equation}
which is the inverse reaction of the dominant LiH dissociation reaction.
Since their rate coefficients are not well contrained,
these reactions have not been included in studies on the first star 
formation.
Considering possible variations of these rates, we study the effect of
different rate coefficients. 
Furthermore, we also assess the effects of different initial conditions, 
which correspond to different sizes of parent clouds, 
while in paper I we only calculated a single case. 
As for the geometry of clouds, we consider not only spherical cores, 
which are assumed in paper I, 
but also filamentary clouds, which are suppposed to form in 
large pre-galactic clouds ($\gsim 10^8\MO$).
As in paper I, we evaluate the luminosities of primordial molecules
both in the accretion and collapse phases, while related works
by Kamaya and Silk (2002,2003) and Ripamonti et al. (2002)
studied only the collapse phase.

Using this model, we discuss the condition, i.e., formation redshift, 
star formation rate (SFR), and size of parent pre-galactic clouds 
of Pop III stars that permit SPICA and JWST to detect primordial 
molecular emission.
We found that those facilities can observe H$_2$ line emission
in a pre-galactic cloud at $z<10$ 
with SFR $\sim 10^{3-4}\MO/{\rm yr}$. 
LiH fraction is found to remain low ($< 10^{-3}$ of total Li nuclei)
even in such high densities where the three-body reaction is 
important.
This is because of the high rate of dissociation reaction
\begin{equation}
{\rm LiH}+{\rm H}\rightarrow {\rm Li}+{\rm H_2}.
\end{equation}

The structure of this paper is as follows: 
In section 2, our calculation method and results 
for spherical star-forming cores are described. 
In section 3, modification in the method and results for 
filamentary clouds are presented.
In section 4, we give discussions about (i) an analytical estimate of 
molecular line luminosities, which agrees well with numerical calculations 
in sections 2 and 3, and (ii) detectability of primordial star-forming regions 
by next-generation facilities, SPICA and JWST. 
Finally, we summarize this paper briefly in section 5.

\section{Evolution of a Spherical Star-Forming Core}
\subsection{The Model}
In this paper, we use the same method as in paper I
in calculating the dynamical and luminosity evolution 
of star-forming cores.
Modification from the previous model is addition of
deuterium and lithium chemical reactions and 
the cooling due to HD and LiH.

\subsubsection{Dynamics}
In our analysis, dynamics in two phases of star formation, i.e., 
the runaway collapse phase and the main accretion phase, 
corresponding to before and after the formation of a central protostar, 
are treated in different ways because of their vastly different 
dynamical natures.

In the run-away collapse phase, the dynamics is described mathematically 
by a Larson-Penston type similarity solution (Larson 1969; Penston 1969). 
In this solution, the cloud can be divided into two regions: 
the central densest region and the lower density envelope.
Since the central densest region collapses at roughly 
the free-fall time, $t_{\rm ff}=\sqrt{3\pi / 32G\rho_{\rm c}}$, 
which inversely depends on the density, the envelope with lower 
density remains almost unevolved.
Thus this evolution is called the run-away collapse.
The central region with a flat density distribution has 
a length scale about the local Jeans length, 
$\lambda_{\rm J}=\sqrt{c_{\rm s}^2\pi/G\rho_{\rm c}}$, where $c_{\rm s}$ and
$\rho_{\rm c}$ are the sound speed and density at the center.
On the other hand, the envelope is characterized by density decreasing
with radius by the power-law $\rho \propto r^{-2/(2-\gamma)}$, where 
$\gamma=d{\rm ln}P/d{\rm ln}\rho$ is the effective adiabatic coefficient 
(Larson 1969).

Here, we model the central region as a sphere with radius
$R_{\rm core}=\alpha \lambda_{\rm J}/2$, 
and collapsing at the free-fall rate with pressure modification: 
\begin{eqnarray}
\frac{d\rho_{\rm c}}{dt}=\frac{\rho_{\rm c}}{\beta t_{\rm ff}}, 
\label{eq:FF}
\end{eqnarray}
where $\alpha$ and $\beta$ are correction factors of order unity 
owing to finite pressure gradient force.
These values will be given below shortly.
Considering the run-away nature of the collapse,  
we assume that physical quantities evolve only when the mass shell 
is inside the central region, i.e. $r<R_{\rm core}$: 
after the matter has left behind the central evolution and become part of 
the envelope, those quantities are kept constant until the end of this phase.
With this assumption, we can construct the whole density and velocity 
structures by calculating only the evolution of central quantities.

The evolution of primordial star-forming clouds in the run-away collapse 
phase is studied by way of one-dimensional (spherical symmetric) 
radiative hydrodynamics by Omukai and Nishi (1998).
They found that actual dynamics is approximated by the 
Larson--Penston type similarity solution (Larson 1969; Penston 1969) 
with $\gamma=1.09$.
For the self-similar solution with $\gamma=1.09$, $\alpha=1.13$ and 
$\beta=2.13$, which we use hereafter.

At the end of the collapse phase, a small protostar forms 
at the center. 
Although the protostar is initially small, it grows in mass by accreting 
the surrounding matter.
This is the onset of the main accretion phase.
In this phase, since the pressure gradient force is negligible compared with 
the gravitational force, we assume that the matter in the envelope infalls 
at the free-fall rate. 
The evolution of each mass shell is given by 
\begin{eqnarray}
\frac{dv}{dt}=-\frac{GM_{r}}{r^2},\\
\frac{dr}{dt}=v,
\end{eqnarray}
where $v$ is the velocity of a free-falling mass shell, 
and $M_{r}$ is the mass contained inside the mass shell. 
The density of a free-falling mass shell, $\rho$, is described by 
$\rho\propto r^{-3/2}$. 
Using these equations, we calculate the evolution of each mass shell 
until the begining of ${\rm H_2}$ dissociation.

\subsubsection{Thermal Evolution}
We follow the evolution of the internal energy per unit mass
\begin{equation}
e=\frac{1}{\gamma_{\rm ad}-1}\frac{kT}{\mu m_{\rm H}}
\end{equation}
by solving the energy equation
\begin{equation}
\frac{de}{dt}=-P\frac{d}{dt}\left(\frac{1}{\rho}\right) - \frac{\Lambda_{\rm tot}}{\rho},
\end{equation}
where 
$P$ is the pressure, $\mu$ is the mean molecular weight, 
and $m_{\rm H}$ is the mass of the hydrogen nucleus. 
The total cooling rate per unit volume, $\Lambda_{\rm tot}$, 
is the sum of the following contributions: 
\begin{equation}
\Lambda_{\rm tot}=\Lambda_{\rm H_2}+\Lambda_{\rm HD}+\Lambda_{\rm LiH}
+\Lambda_{\rm chem},
\end{equation}
where $\Lambda_{\rm H_2}$ is the cooling rate by ${\rm H_2}$ line emission, 
$\Lambda_{\rm HD}$ by HD line emission and $\Lambda_{\rm LiH}$ 
by LiH line emission, and $\Lambda_{\rm chem}$ is 
the net cooling rate by chemical reactions. 
The dominant contribution to $\Lambda_{\rm chem}$ is the cooling by 
H$_2$ dissociation. 
Although we have also included the cooling by continuum emission 
as in paper I, it is always negligible in comparison with other terms.
The H$_2$ cooling is treated in the same way as in paper I, where
we considered effects of radiation trapping for optically thick lines 
by the escape probability method.
In addition to the above processes considered in paper I,
here we include the HD and LiH cooling. 
For the collisional de-excitation rate of HD, we use the rate 
coefficients of Flower et al. (2000). 
We take account of the first eight rotational levels of HD.
In calculating the emissivity of HD lines, 
we use the same method as for ${\rm H_2}$ lines.
The cooling rate by LiH lines $\Lambda_{\rm LiH}$ is given by
\begin{equation}
\Lambda_{\rm LiH}=\frac{\Lambda_{\rm LiH}({\rm LTE})}
{1+n_{\rm cr}/n({\rm H})}.
\end{equation}
The critical density for LTE $n_{\rm cr}$ is
\begin{equation}
n_{\rm cr}/n({\rm H})=\frac{\Lambda_{\rm LiH}({\rm LTE})}
{\Lambda_{\rm LiH}[n({\rm H}) \rightarrow 0]}, 
\end{equation}
where
$\Lambda_{\rm LiH}({\rm LTE})$ is the cooling rate at the LTE,
$\Lambda_{\rm LiH}[n({\rm H}) \rightarrow 0]$ is that in the low density
limit.
For $\Lambda_{\rm LiH}[n({\rm H}) \rightarrow 0]$, 
we use the fit by Galliand Palla (1998).

\subsubsection{Chemistry}
In addition to the hydrogen chemical network in paper I, we solve 
major deuterium chemistry between D, ${\rm D^+}$, ${\rm D^-}$, HD, and  
${\rm HD^+}$, and lithium chemistry between 
Li, ${\rm Li^+}$, ${\rm Li^-}$, LiH, ${\rm LiH^+}$.
The included D reactions and their coefficients are the same as
in Nakamura and Umemura (2002; Table 1). 
The Li reactions include those in Stancil, Lepp, and Dalgarno (1996) 
and are given in Table 1. 

\begin{table}
\begin{center}
\begin{tabular}{c l l}
\multicolumn{3}{c}{Table 1. Additional Li Chemical Reactions.}\\ \hline\hline
Numbers &Reactions  & Rate Coefficients  \\ \hline
1 &Li + H$_2$ $\rightarrow$ LiH + H 
&$k_{1}=2.0\times 10^{-11}(Q_{\rm LiH}/Q_{\rm H_2})
{\rm exp}(-2.28 \times 10^4/ T)~{\rm cm^3}$ $\dagger$ \\
2 &Li + 2H $\rightarrow$ LiH + H
&$k_{2}=2.5\times 10^{-29}/T~{\rm cm^6}$ \\
3 &Li + H + H$_2$ $\rightarrow$ LiH + H$_2$
&$k_{3}=4.1\times 10^{-30}/T~{\rm cm^6}$ \\
\hline
\end{tabular}
\end{center}
\caption{In addition to the Li reactions except photoreactions 
in Table 1 of Stancil, Lepp, and Dalgarno (1996), 
we included the three reactions above.\\
$\dagger$ $Q_{\rm H_2}$ and $Q_{\rm LiH}$ are 
partition functions of H$_2$ and LiH from Sauval and Tatum (1984). }
\end{table}
Stancil, Lepp, and Dalgarno (1996) discussed possible 
enhancement of LiH via the three-body reactions
\begin{eqnarray}
{\rm Li + H + H \rightarrow LiH + H}, \\
{\rm Li + H + H_2 \rightarrow LiH+ H_2},
\end{eqnarray}
at high enough densities.
We also include these reactions.
Without accurate measuremnt of the rate coefficients for these reactions, 
we estimate them by assuming the same reaction cross sections as those 
of the corresponding hydrogen three-body reactions (Palla et al. 1983) 
but with correction owing to different molecular weight.
The dissociative reaction 
\begin{equation}
{\rm LiH + H \rightarrow Li + H_2} 
\label{eq:Lidiss}
\end{equation}
is important for limiting LiH fraction owing to 
its high reaction coefficient 
$k_{\rm diss}=2.0 \times 10^{-11} {\rm cm^{3}}$.
We include the inverse reaction
\begin{equation}
{\rm Li + H_2 \rightarrow LiH + H} 
\label{eq:Limolex}
\end{equation}
with its rate coefficient calculated from the detailed balance.

\subsubsection{Luminosity Calculation}
For the thermal evolution, we employ an assumption similar to that 
for the dynamics:
physical quantities evolve only when the mass element is inside the 
central region.
After the mass element is left behind the central collapse at some instance, 
the physical quantities of the matter are kept constant 
until the end of the collapse phase. 
In this way, we can construct the distribution of temperature and 
chemical compositions in the entire cloud, as well as 
the density and velocity structures.
Using this cloud structure, we calculate the molecular line emissivity
in each part of the cloud (see 2.1.2) and estimate the total line luminosity 
by summing those from the central region and from the envelope.

\subsection{Initial conditions}
\begin{table}
\begin{center}
\begin{tabular}{c c c}
\multicolumn{3}{c}{Table 2. Initial model parameters.}\\ \hline\hline
model  &$T({\rm K})$ &$y({\rm H_2})$ \\ \hline
S1 (fiducial) &200  &$5\times 10^{-4}$  \\
S2  &100  &$5\times 10^{-4}$  \\
S3  &300  &$5\times 10^{-4}$  \\
S4  &400  &$5\times 10^{-4}$  \\
S5  &200  &$2\times 10^{-3}$  \\
S6  &200  &$10^{-3}$  \\
S7  &200  &$10^{-4}$  \\
\hline
\end{tabular}
\end{center}
\caption{The initial condition is taken at the number density 
$10^{4}{\rm cm^{-3}}$}
\end{table}
We start calculations at the number density of $10^{4}{\rm cm^{-3}}$.
We calculate seven models whose initial temperature and 
H$_2$ concentrations are given in Table 2. 
The model S1, which we call the fiducial model hereafter, represents 
the typical parameter set of first-star forming cores 
(e.g., Bromm et al. 2002).
Around the fiducial parameters, we put possible variations of initial 
temperature and ${\rm H_2}$ abundance for other models.

In particular, the difference in ${\rm H_2}$ abundance 
reflects the past history of star-forming cores in the parent cloud.
If mass of the parent cloud is $\sim 10^6 \MO$, i.e., 
at the size of the first luminous objects in the standard 
${\rm \Lambda}$CDM cosmology, the ${\rm H_2}$ abundance 
reaches at most $y({\rm H_2})\sim 10^{-4}-10^{-3}$ at the formation of 
star-forming core ($\sim 10^{4}{\rm cm^{-3}}$; e.g., Abel et al. 2002; 
Bromm et al. 2002).
On the other hand, if the parent cloud is a giant pre-galactic cloud, 
i.e., $\gsim10^8 \MO$, and  its virial temperature is higher than $10^4{\rm K}$, 
the abundant ${\rm H_2}$ is formed ($y({\rm H_2}) \sim 10^{-3}-10^{-2}$) 
owing to shock ionization and subsequent non-equilibrium H$_2$ formation 
(Shapiro, Kang 1987; Susa et al. 1998; Oh, Haiman 2002).
Considering the difference in parent cloud mass, we calculate models of both   
low- (models S1, S7) and high- (models S5, S6) ${\rm H_2}$ abundance. 
Similarly, for observing dependence on different initial temperatures 
from the fiducial one, we calculate models S2-4.
The ionization degree is set to be $y({\rm e^-})=10^{-8}$.
The abundances of other chemical species 
are fixed to the values calculated for the study of molecule formation 
in the primordial gas by Galli and Palla (1998): 
$y({\rm D})=4\times 10^{-5}$, 
$y({\rm HD})=1.2\times 10^{-9}$, $y({\rm Li})=10^{-10}$, 
$y({\rm Li^+})=1.2\times 10^{-10}$, $y({\rm LiH})=7.1\times 10^{-20}$, 
$y({\rm LiH^+})=9.4\times 10^{-18}$. 
The initial abundance of ${\rm H_2^+}$, ${\rm H_3^+}$, ${\rm H^-}$, 
${\rm D^+}$, ${\rm D^-}$, ${\rm HD^+}$, ${\rm Li^-}$ 
are set to zero.
The concentration of the $i$th species $y(i)$ is defined by 
$y(i)=n(i)/n_{\rm H}$, where $n(i)$ is the number density of
the $i$th species and $n_{\rm H}$ is the number density 
of hydrogen nuclei.

\subsection{Numerical Results}
\subsubsection{Evolution in the Fiducial Case}

\begin{figure}
  \begin{center}
    \FigureFile(80mm,80mm){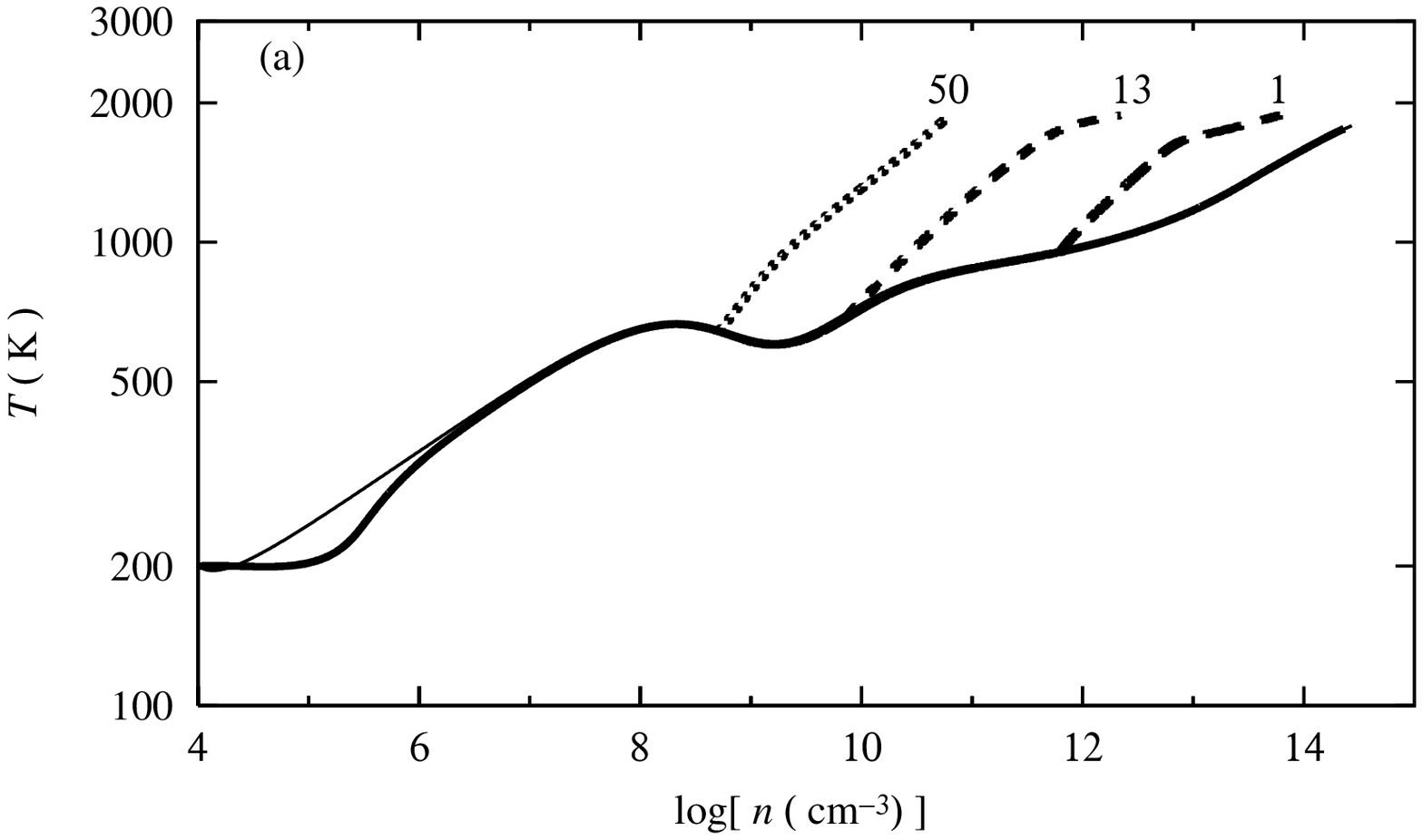}
    \FigureFile(80mm,80mm){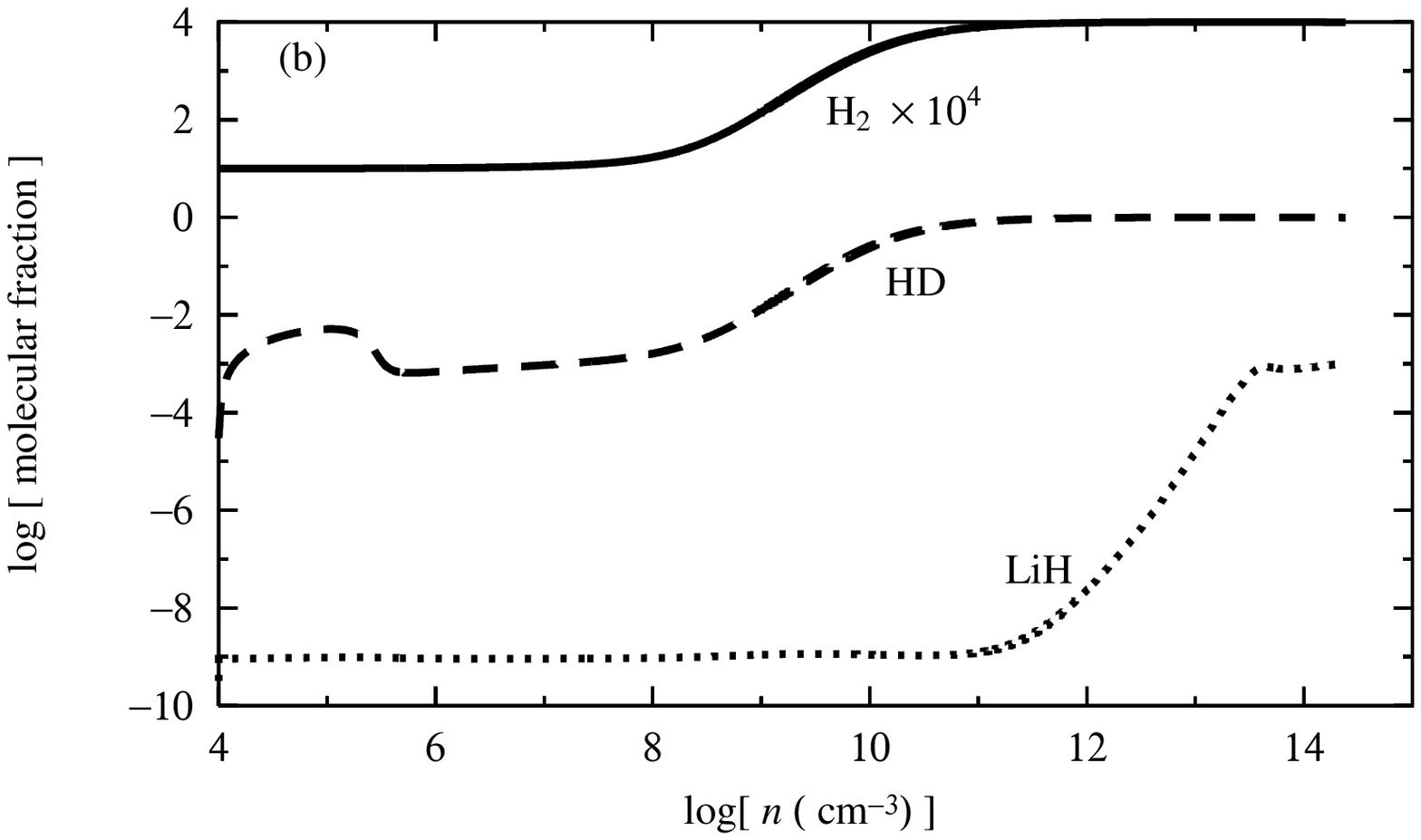}
  \end{center}
\caption{Evolution of (a) the temperature and  
(b) the molecular fractions of H (solid) and D (dashed) and 
Li (dotted) against the number density for the fiducial 
model (S1). 
The molecular fraction is the fraction of each element  
locked into the molecular species shown.
In (a), the temperature in the central part in the run-away collapse 
phase is drawn with a thick solid line. 
Here, that of the model with the same parameters as 
in the fiducial model but without HD cooling (thin line) is also shown
for comparison.
For the accretion phase, the evolutionary trajectories of the mass 
shells of $M/ \MO = $ 1 (long-dashed lines), 
13 (short-dashed lines) and 50 (dotted lines) are shown.
In (b), only the evolutionary trajectories of the central region 
are shown. 
The H$_2$ fraction is shifted by $10^{4}$ for clarity.
}
\label{fig:TH2}
\end{figure}

The evolution of the star-forming core in the fiducial case 
(model S1) is presented in Figure \ref{fig:TH2}, where (a) the temperature 
and (b) the ${\rm H_2}$, HD, and LiH fractions are shown 
against the number density. 
Here, the molecular fraction denotes the fraction of 
each element converted into the molecular species. 
For example, for LiH, the molecular fraction $f({\rm LiH})$ 
is defined as $f({\rm LiH})=n({\rm LiH})/n_{\rm Li}$, 
where $n_{\rm Li}$ is the number density of Li nuclei. 
In Figure \ref{fig:TH2}(a), the solid curves show the evolution of 
the central region in the collapse phase. 
For the accretion phase, the evolution of three representative 
mass-shells is indicated by long- and short-dashed and dotted lines.  
The bold and thin lines indicate the temperature 
of the fiducial model and of the model with same parameters 
but without HD cooling, respectively.
In Figure \ref{fig:TH2}(b), the H$_2$, HD and LiH fractions at 
the central region in the collapse phase are shown by 
solid and dashed and dotted lines. 

The small enhancement of HD fraction and resultant HD cooling 
in $n<10^5{\rm cm^{-3}}$ (Figure \ref{fig:TH2}b) causes a slight deviation 
in temperature evolution from the case without HD 
(see Figure \ref{fig:TH2}a). 
With increasing temperature for $n>10^5{\rm cm^{-3}}$, 
this HD is soon dissociated. 
The reason for this sudden HD dissociation is as follows:
the abundance of HD in this density range is determined 
by balance between formation via  
\begin{equation}
{\rm D^+}+{\rm H_2} \rightarrow {\rm H^+}+{\rm HD}, 
\end{equation} 
and dissociation via
\begin{equation}
{\rm HD}+{\rm H^+} \rightarrow {\rm H_2}+{\rm D^+}. 
\end{equation}
The equilibrium abundance between these reactions is given by 
$y({\rm HD})/y({\rm H_2})
=2{\rm exp}( 464{\rm K}/T ) y({\rm D^+})/y({\rm H^+})$, 
which means that for low temperatures ($T < 464$K),  
HD fraction is enhanced relative to H$_2$ fraction. 
When the density exceeds the critical density 
$n_{\rm cr} \sim 10^5 {\rm cm^{-3}}$ for HD to reach the 
local thermodynamic equilibrium, the temperature begins to increase.
Since the increase in temperature reduces the HD fraction, 
HD cooling becomes inefficient.
At higher density, the two evolutionary paths in Figure \ref{fig:TH2}(a) 
converge.
Accordingly, the presence of HD does not alter the evolution 
in the fiducial case for higher densities.
In particular, the HD cooling does not modify the temperature evolution 
of mass shells in the accretion phase (Figure \ref{fig:TH2}a).

\begin{figure}
  \begin{center}
    \FigureFile(80mm,80mm){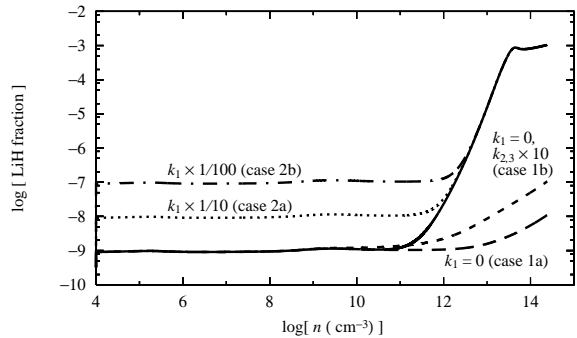}
  \end{center}
\caption{ 
Evolution of LiH abundance for different formation rate coefficients.
Shown are the values at the center in the collapse phase. 
The standard case, whose rates are given in Table 1, is shown by the
solid line. 
The cases 1a, b, where the molecular exchange reaction 
(reaction 1 in Table 1) are omitted, are shown 
by long-dashed (1a) and short-dashed lines (1b).
In the case 1a the coefficients of the three-body reactions
(reactions 2,3 in Table 1) are the same as the standard values, 
while in the case 1b they are increased by a factor of ten. 
The cases with reduced rates of the molecular exchange reaction 
and its inverse by a factor of 10 (case 2a) or 100 (2b) from 
the standard ones are shown by dotted and long-dash-dotted lines. 
 }
\label{fig:LiH}
\end{figure}

Hydrogen becomes fully molecular at $\sim 10^{11} {\rm cm^{-3}}$
via the three-body reactions, 
\begin{eqnarray}
3{\rm H}\rightarrow {\rm H_2}+{\rm H},\\
2{\rm H}+{\rm H_2}\rightarrow 2{\rm H_2},
\end{eqnarray} 
which start at $\sim 10^8 {\rm cm^{-3}}$ (Palla et al. 1983).
Simultaneously, all the deuterium becomes HD via the molecular 
exchange reaction
\begin{equation}
{\rm D + H_2 \rightarrow H + HD}.
\end{equation} 
On the other hand, Figure \ref{fig:LiH} shows 
the effect of different LiH formation rate
coefficients on the LiH fraction.
Along with the case with the standard rate given in Table 1,
following cases are shown:

[Case 1] the molecular exchange reaction 
(reaction 1 in table 1) is excluded from chemistry.
The rate of three-body reactions (reactions 2,3 in Table 1) 
are the same as the standard values (case 1a) or is increased 
by a factor of ten (1b).

\begin{figure}
  \begin{center}
    \FigureFile(80mm,80mm){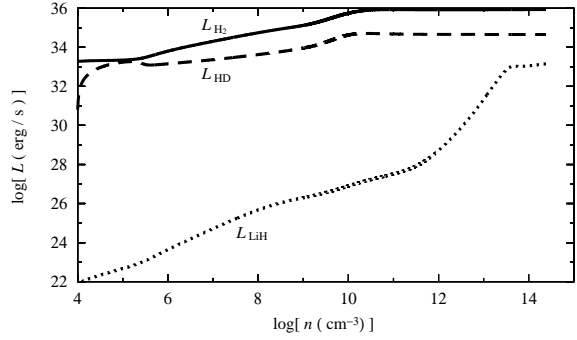}
  \end{center}
\caption{ 
Evolution of the cooling rate by H$_2$ (solid line), HD (dashed line) 
and LiH (dotted line) in the run-away collapse phase 
against the number density for the fiducial model (S1).
 }
\label{fig:Lm}
\end{figure}

[Case 2] the rates of the molecular exchange reaction and 
its inverse are reduced by a factor of 10 (case 2a) or 100 (2b) 
from the standard ones. 
In Figure \ref{fig:LiH}, the LiH fraction in $n<10^{12}{\rm cm^{-3}}$
increases if the rate coeffecients of the reaction 1 and 
its inverse are decreased (i.e, case 2): 
since the LiH formation is suppressed  by the dissociative reaction 
(equation \ref{eq:Lidiss}; Stancil et al. 1996; Bougleux, Galli 1997; 
Puy, Signore 1998),
lowering the rate coeffecient for this reaction results in 
enhancement of the LiH fraction.
However, even in these cases, the LiH fraction remains $<10^{-7}$ for 
$n<10^{12}{\rm cm^{-3}}$ and 
their cooling rate is always negligible.
At $\sim 10^{12} {\rm cm^{-3}}$, the LiH fraction increases
with the onset of the molecular exchange reaction (reaction 1 in Table 1).
The Li three-body reactions are negligible in comparison with
the exchange reaction for LiH formation.
For  higher density, the molecular fraction of lithium saturates at the
equilibrium value $\sim 10^{-3}$
(Figure \ref{fig:TH2}b, \ref{fig:LiH}).
As in Figure \ref{fig:Lm}, the total cooling rate by LiH $L_{\rm LiH}$ is
less than $10^{28} {\rm erg~s^{-1}}$ in $n<10^{12}{\rm cm^{-3}}$.
At  $n \sim 10^{13}{\rm cm^{-3}}$, 
the maximum value $L_{\rm LiH} \sim 10^{33} {\rm erg~s^{-1}}$ is reached, 
which is lower than $L_{\rm H_2}$ and $L_{\rm HD}$ by 2-3 orders of
magnitude.
Some authors (Lepp, Shull 1984; Stancil et al. 1996) considered that
in some late stages of the collapse all the lithium might 
be converted to LiH by the three-body reactions, 
and thus LiH could be an important coolant.
However, in our calculation, LiH never become an important coolant in any
evolutionary stage.
As for HD, $L_{\rm HD}$ is smaller by more than an order
of magnitude than $L_{\rm H_2}$ for $n>10^6{\rm cm^{-3}}$,
although $L_{\rm HD}$ is comparable to $L_{\rm H_2}$
at $n<10^6{\rm cm^{-3}}$ in the fiducial case (Figure \ref{fig:Lm}).

\subsubsection{Dependence on Initial Conditions}

\begin{figure}

  \begin{center}
    \FigureFile(80mm,80mm){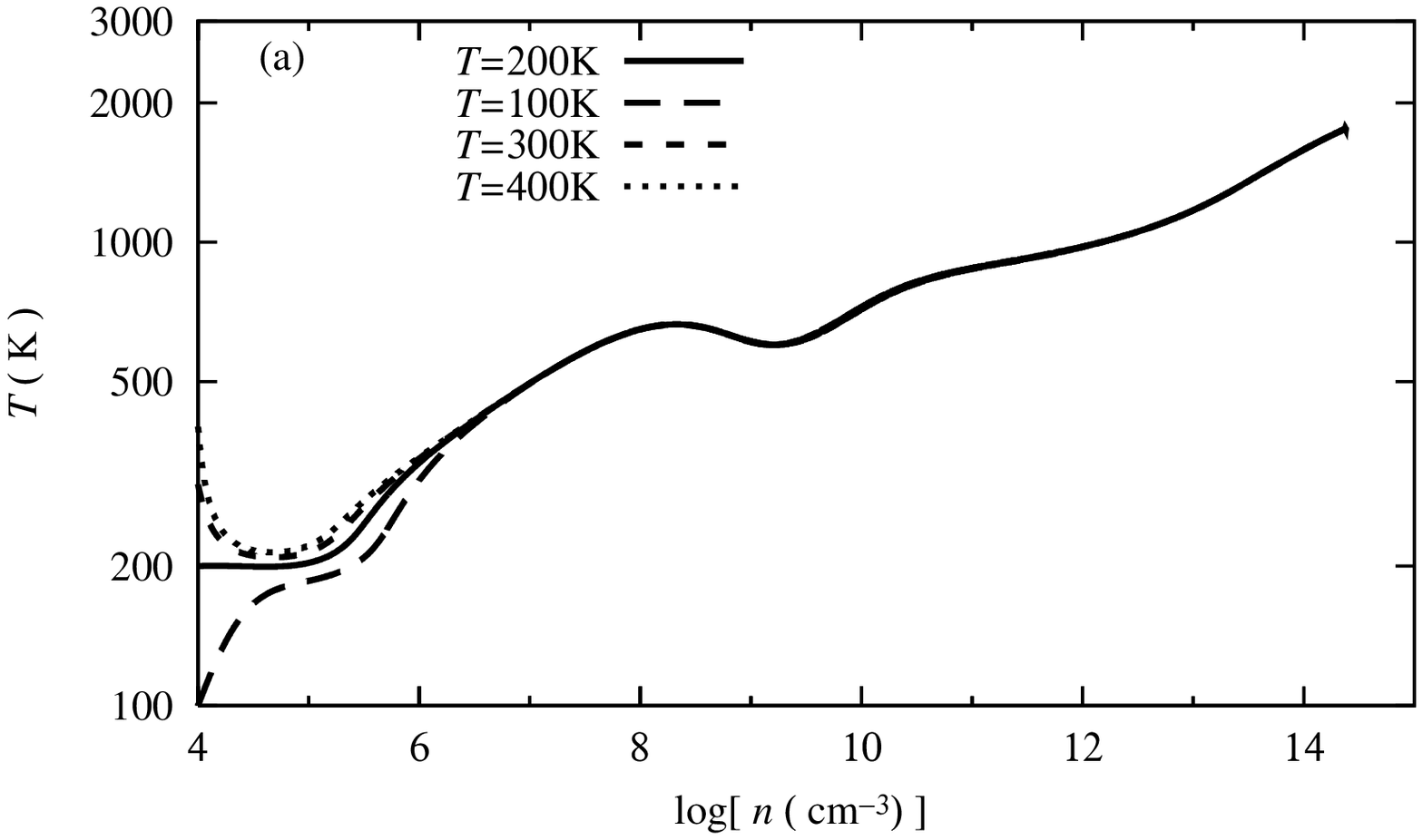}
    \FigureFile(80mm,80mm){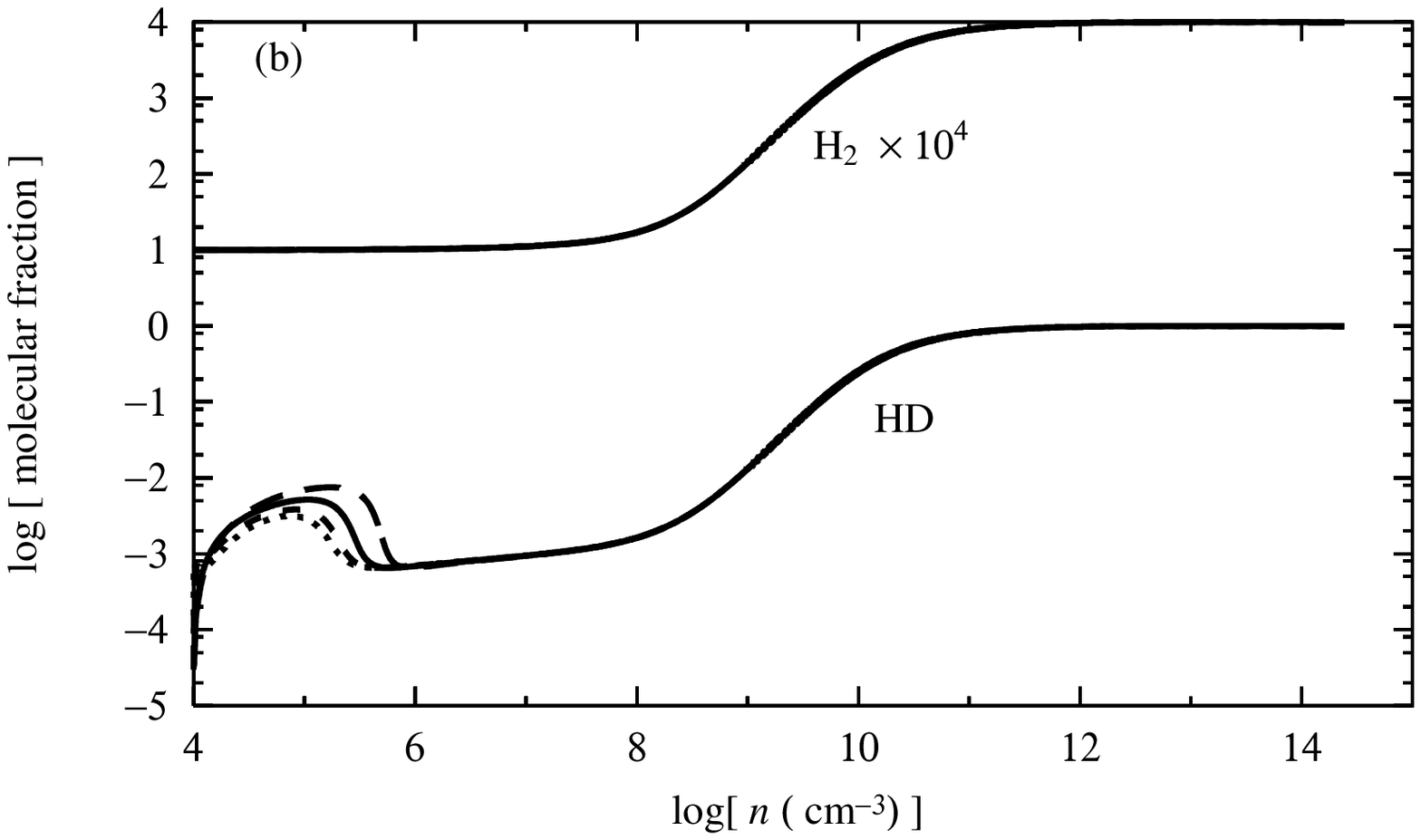}
  \end{center}
\caption{Effect of difference at the initial temperature $T_{\rm init}$: 
(a) temperature evolution and 
(b) evolution of the H$_2$ (upper lines) and HD (bottom lines) fractions 
against the number density for the run-away collapse phase. 
The evolutionary trajectories of model S1 (fiducial): 
$T_{\rm init}=200$K (solid), 
S2: $T_{\rm init}=100$K (long-dashed), 
S3: $T_{\rm init}=300$K (short-dashed) and 
S4: $T_{\rm init}=400$K (dotted) are shown.}
\label{fig:TH2_ABCD}
\end{figure}

In the following, we describe the dependence on initial conditions. 
First, we show the dependence on initial temperature
by comparing the fiducial model with models S2-4 
(Figure \ref{fig:TH2_ABCD}).
For the initial temperature $\gsim 200$K, temperature drops,
since the H$_2$ cooling is effective in this temperature range.
On the other hand, for lower initial temperatures, the temperature 
rises with the compressional heating owing to inefficient cooling.
Then, the temperature approaches around $T \simeq 200$K, and 
for $n\sim 10^5{\rm cm^{-3}}$, these evolutionary trajectories 
converge in all these cases.

\begin{figure}
  \begin{center}
    \FigureFile(80mm,80mm){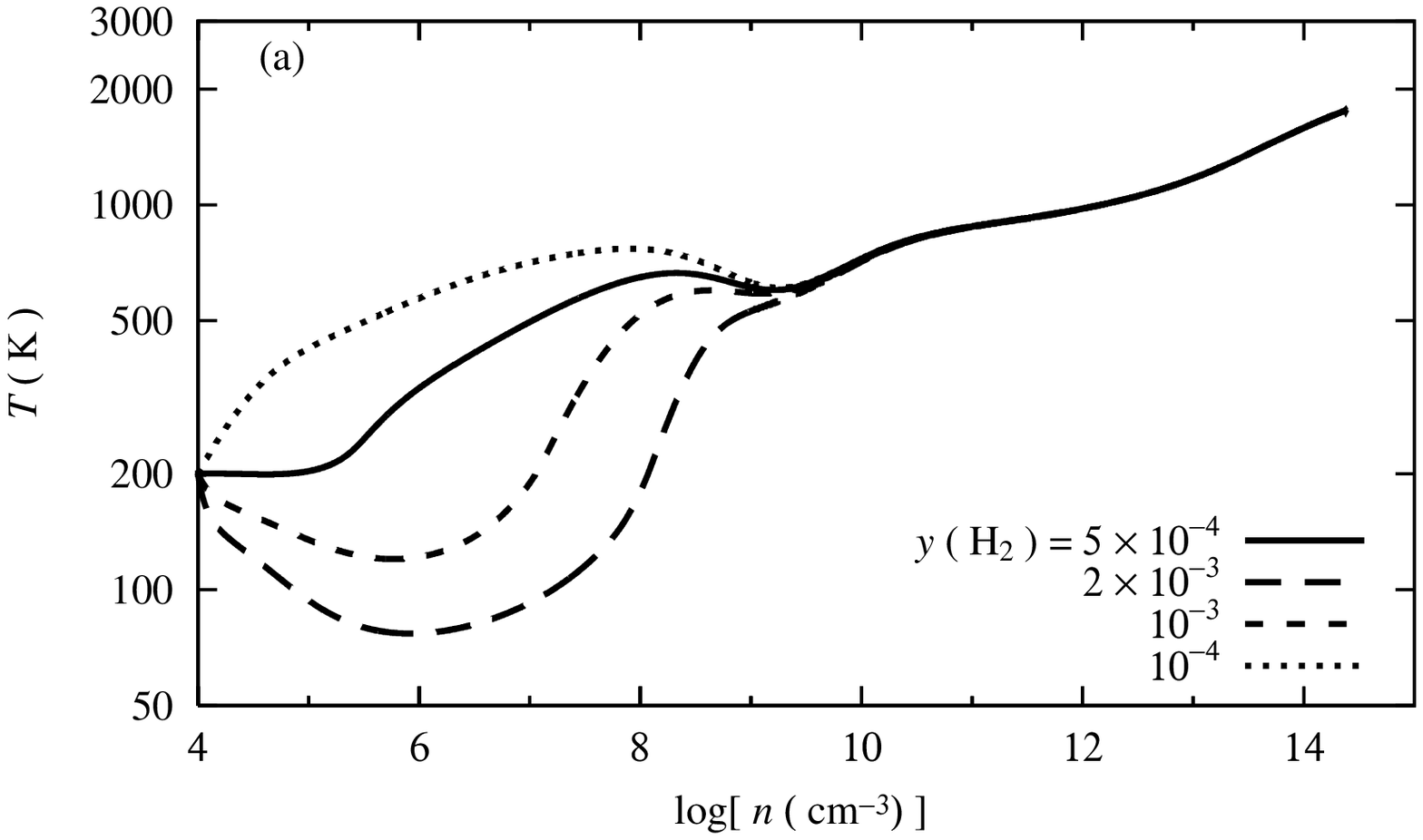}
    \FigureFile(80mm,80mm){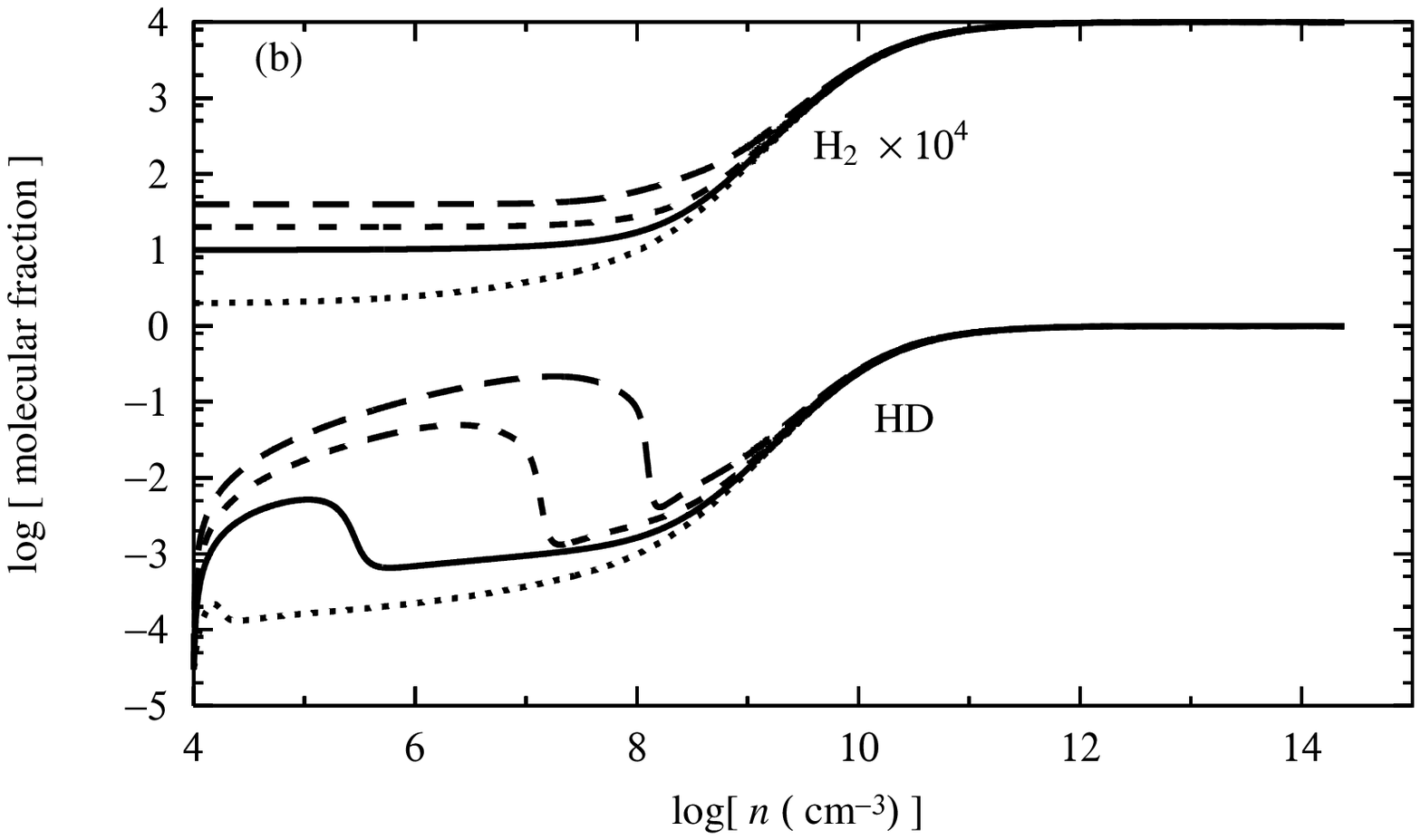}
  \end{center}
\caption{Effect of difference in the initial H$_2$ concentration 
$y_{\rm init}({\rm H_2})$.
Meaning of the panels is the same as in Figure 2.
The evolutionary trajectories of model S1 (fiducial): 
$y_{\rm init}({\rm H_2})=5\times 10^{-4}$ (solid), 
S5: $y_{\rm init}({\rm H_2})=2\times 10^{-3}$ (long-dashed), 
S6: $y_{\rm init}({\rm H_2})=10^{-3}$ (short-dashed) and 
S7: $y_{\rm init}({\rm H_2})=10^{-4}$ (dotted) are shown.}
\label{fig:TH2_AEFG}
\end{figure}

Next, we see the dependence on initial ${\rm H_2}$ abundance 
$y_{\rm i}({\rm H_2})$ by comparing between the fiducial model and 
models S5-7 (Figure \ref{fig:TH2_AEFG}). 
For low initial H$_2$ abundances $y_{\rm i}({\rm H_2}) \lsim 10^{-3}$,
the temperature does not drop sufficiently for HD formation.
In this case, without enough HD formation, 
the temperature evolution follows the fiducial case closely
(see Figure \ref{fig:TH2_AEFG}a).
The situation changes for higher values of 
initial H$_2$ abundance $y_{\rm i}({\rm H_2})\geq 10^{-3}$, 
(Uehara, Inutsuka 2000; Nakamura, Umemura 2002).
With such a high value of  $y_{\rm i}({\rm H_2})$, 
the temperature falls below the threshold value 
($\sim 150$K) for HD formation.  
Because of the rapid HD formation and consequent HD cooling, 
the temperature drops further. 
Although HD cooling becomes inefficient after the density exceeds the critical
value for LTE ($\sim 10^{5}{\rm cm^{-3}}$), the temperature evolution differs 
from the fiducial case owing to different H$_2$ abundance.
When the difference in H$_2$ abundance becomes negligible  
due to the three-body reactions
at $\sim 10^{9}{\rm cm^{-3}}$, the temperature evolution converges. 
In above cases with high initial H$_2$ abundance (models S5 and S6),
the temperature rises adiabatically around $n=10^{6}-10^{8}{\rm cm^{-3}}$.
In this regime, the actual contraction of clouds should be slower than
the assumed free-fall rate.
Longer contraction timescale reduces the compressional heating rate, and
thus the temperature increase should be more gradual than in our model.
Nevertheless, alteration of the evolution at this density range
does not affect the maximum luminosity:
in such high density regime where the maximum luminosity is attained, 
the evolution has converged already regardless of the behavior in 
the lower density regime. 

In conclusion, the evolutionary trajectories of both temperature and 
HD abundance are most affected by the initial $y({\rm H_2})$ value.
This difference disappears $n\sim 10^9{\rm cm^{-3}}$ 
with rapid H$_2$ formation by the three-body reactions.
The evolution at high densities is independent of the initial conditions.
In particular, the evolution of inner mass shells ($M_{r}\lesssim 50\MO$) 
is independent of the initial conditions because the density there is 
higher than $10^9{\rm cm^{-3}}$.

\subsubsection{Luminosity Evolution}

The maximum luminosities $L_{\rm peak}$ of the two strongest H$_2$ lines 
1--0 Q(1) and 0--0 S(3)
\footnote
{For example, the notation 1--0 Q(1) means the following transition: 
`` 1--0 " denotes vibrational transition from $v=1$ to 0. 
Q denotes the change of rotational quantum number $\Delta J=0$.
Similarly, O, P, R, and S, denote $\Delta J=2, 1, -1$, and -2, respectively.
The number in brackets denotes the rotational quantum number 
after the transition.},
and the mass of the protostar when these maximum luminosities are 
attained, $M_{\rm peak}$ are presented in Table 3 for each model.
\begin{center}
\begin{tabular}{c c c c c}
\multicolumn{5}{c}{Table 3. Maximum Line Luminosities.}\\ \hline\hline
model  & $L_{\rm peak}$(1-0 Q(1))($10^{35}$erg ${\rm s^{-1}}$) 
&$M_{\rm peak}(\MO)$  
&$L_{\rm peak}$(0-0 S(3))($10^{35}$erg ${\rm s^{-1}}$) 
&$M_{\rm peak}(\MO)$ \\ \hline
S1  &1.69  &12.9  &1.11  &18.2  \\
S2  &1.70  &13.8  &1.11  &18.2  \\
S3  &1.69  &13.8  &1.11  &18.2  \\
S4  &1.69  &13.8  &1.11  &18.2  \\
S5  &1.50  &12.0  &0.70  &7.9   \\
S6  &1.63  &12.9  &0.91  &10.5  \\
S7  &1.94  &29.5  &1.33  &19.5  \\
\hline
\end{tabular}
\end{center}

\begin{figure}
  \begin{center}
    \FigureFile(110mm,80mm){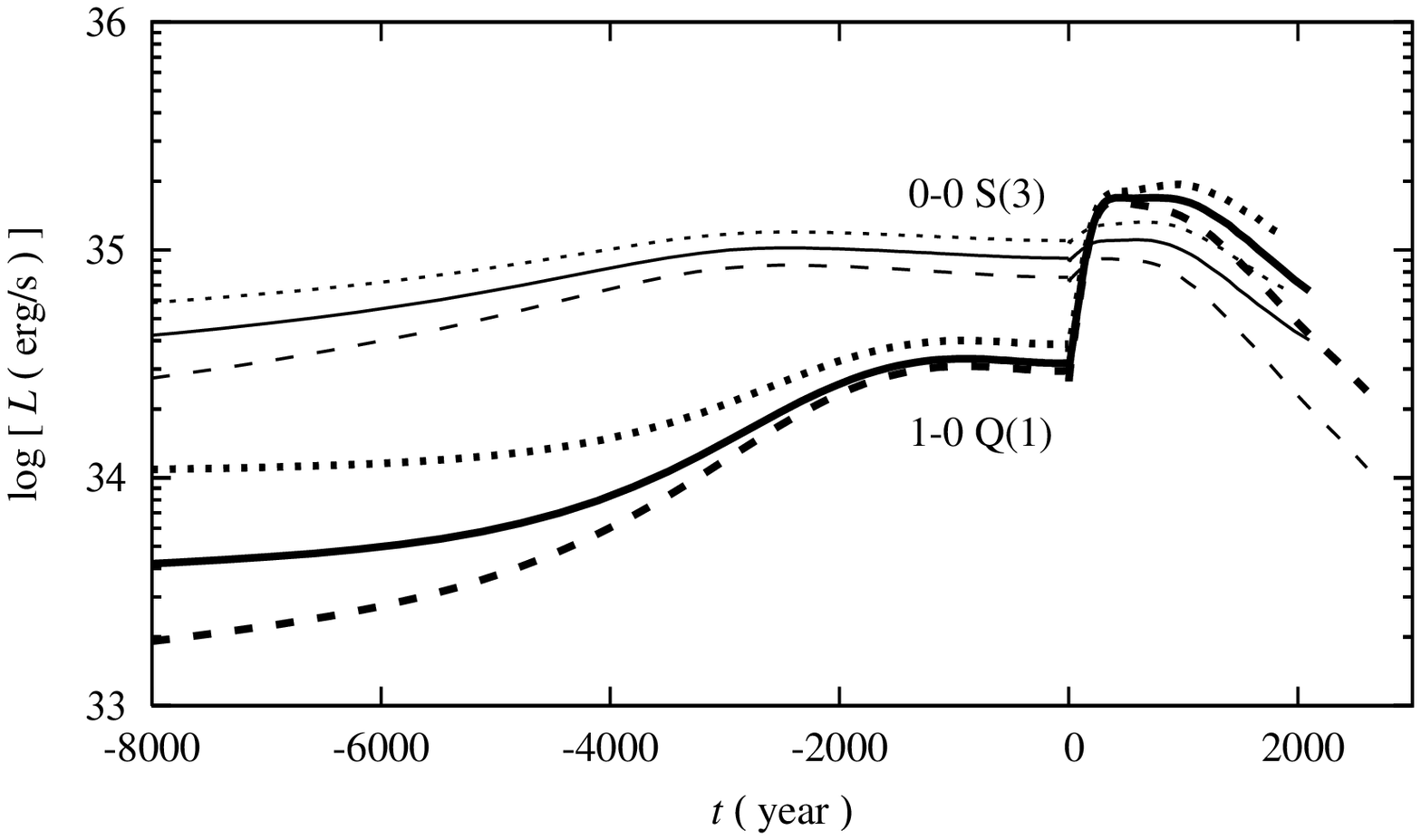}
    \FigureFile(110mm,80mm){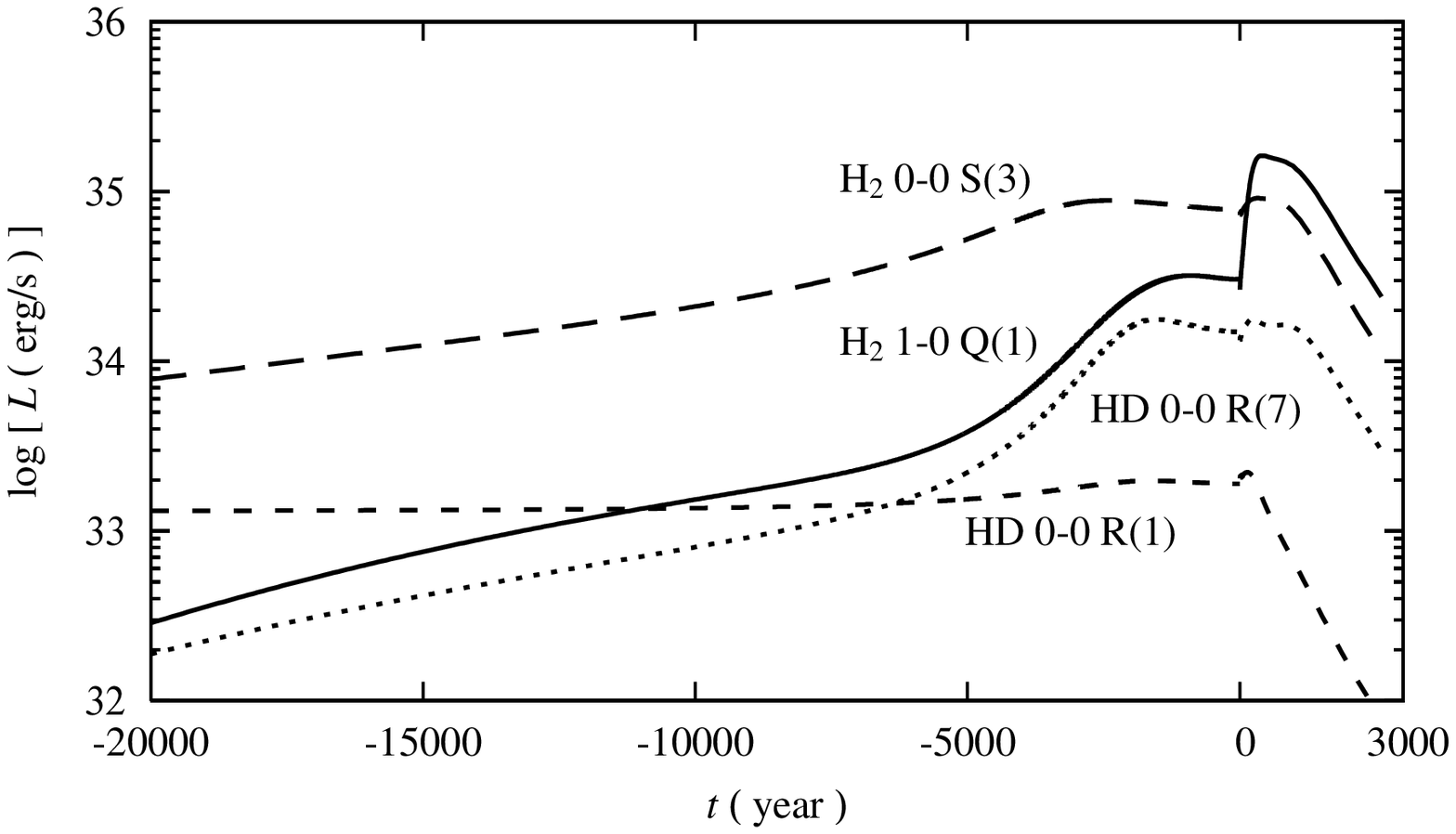}
  \end{center}
\caption{
Luminosity evolution of strong H$_2$ and HD lines. 
(a) Those of the two H$_2$ liens, 1--0 Q(1) (thick lines) and 
0--0S(3) (thin lines), are shown for models S1 (fiducial; solid), 
S6 (dashed) and S7 (dotted). 
(b) Those of the same H$_2$ lines (solid and long-dashed lines) as (a), 
and of the two HD lines, 0--0 R(1) (short-dashed) and 0--0 R(7) (dotted) 
are shown for model S6. 
Formation of the central protostar corresponds to $t=0$. 
Evolution up to the protostar mass of $70\MO$ is shown. 
} 
\label{fig:L(time)sp}
\end{figure}

As examples, we choose the two strongest ${\rm H_2}$ lines 
in our calculation, the rovibrational line of 1--0 Q(1) 
and pure rotational line of 0--0 S(3), whose wavelengths at rest 
are $2.34{\rm \mu m}$ and $10.03{\rm \mu m}$, respectively. 
For models S1, S2, S3, and S4 (cases with same initial H$_2$ abundances 
but different initial temperatures), 
the luminosities evolve in a similar way, 
since both the temperature and HD abundance converge immediately.
In Figure \ref{fig:L(time)sp}(a),
 we show the luminosity evolution of models S1, S6, and S7, which are
cases with the same initial temperatures, but with different initial H$_2$ 
abundances.
The temperature, ${\rm H_2}$ and HD abundances in these models 
are significantly different up to $10^9{\rm cm^{-3}}$.
For lower densities, the luminosity from the cores depends on initial 
conditions (Figure \ref{fig:L(time)sp}a), while 
for higher densities ($>10^9{\rm cm^{-3}}$) physical quantities converge. 
Since temperature is higher in this range than in lower densities and  
the emission from the highest temperature region dominates 
the total emission (see \S 4.1 later),  the similar temperatures
lead to convergence of H$_2$ line luminosities 
for both the late collapse phase and the early accretion phase. 
Note that regardless of initial conditions, 
the peak values of the rovibrational line luminosities are almost unchanged: 
among our models, variations in $y_{\rm i}({\rm H_2})$ cause 
difference in the peak values by less than a factor of two.
In Figure \ref{fig:L(time)sp}(b), we show the luminosity evolution 
of the HD lines 0--0 R(1) ($56.2 {\rm \mu m}$)
and 0--0 R(7) ($15.0{\rm \mu m}$), and that of the H$_2$ lines 
1--0 Q(1) and 0--0S(3) for the model S6.
This is the case with the same parameters as in paper I.
Although the main coolant is still H$_2$ as in paper I, 
its cooling rate is reduced because of the contribution 
by HD cooling for the early collapse phase. 
Since physical quantities converge 
for higher densities ($>10^9{\rm cm^{-3}}$) 
regardless of HD contribution in low density region 
(Figure \ref{fig:TH2}, \ref{fig:TH2_AEFG}), 
the maximum luminosities of H$_2$ lines (Table 1 in paper I) 
remain almost the same.

\section{Evolution of a Filamentary Cloud}
A large pre-galactic cloud ($\gsim 10^8\MO$) collapses into 
a sheet-like configuration because of an instability against 
non-spherical perturbation (Lin et al. 1965; Hutchins 1976; 
Susa et al. 1996).
This oblate cloud experiences a shock and forms a postshock-cooled layer. 
The cooled layer more likely fragments into filamentary clumps 
than into spherical ones (Miyama et al. 1987a, 1987b). 
In the shocked layer, formation of H$_2$ and HD is promoted 
because of abundant free electrons produced by the shock heating, 
which catalyze the formation reactions
(MacLow, Shull 1986; Shapiro, Kang 1987). 
The filaments continue to contract due to effective molecular 
cooling.
At $\sim 10^{9}{\rm cm^{-3}}$, where heat input by H$_2$ three-body 
formation becomes significant, 
the filaments fragment and dense star-forming cores are produced
(Nakamura, Umemura 2002).
This is in contrast to the situation in a small cloud ($\sim 10^6\MO$), 
where cores directly form at $\sim 10^{4}{\rm cm^{-3}}$ 
(Bromm et al. 1999,2002; Abel et al. 2000,2002).
Because of a larger molecular abundance, stronger molecular emission is 
expected in filamentary clouds. 
Here, we then evaluate the molecular line luminosity from those filaments.

\subsection{Model}
In studying the thermal and chemical evolution of a filamentary cloud, 
we approximate the filament by an infinitely long cylindrical cloud, 
for simplicity.
Since a sheet fragments when pressure retards gravitational
contraction, in filaments thus formed, the pressure is comparable 
to the gravity.
Moreover, the ratio of gravity to pressure force is constant during 
the isothermal evolution, which is indeed approximately the case.
The small deviation from the isothermal evolution has an important 
dynamical effect. 
Here, we need to treat the effect of gas pressure carefully 
in our model. 

A contracting filament consists of two parts, 
a denser spindle with almost flat density at the center 
and an envelope with lower denisty (e.g., Inutsuka, Miyama 1997). 
As in the case of spherical cores (section 2), 
we focus on this central dense spindle, and apply the virial equation for 
a cylinder with uniform density:
\begin{eqnarray}
\frac{1}{2}\frac{d^2 {\cal I}}{dt^2}=2{\cal T}+2\Pi-Gm^2,
\label{eq:virial}
\end{eqnarray}
where
\begin{eqnarray}
{\cal I}=\int_{V}\rho r^2 dV,~{\cal T}=\int_{V}\frac{1}{2}\rho u^2dV,
~\Pi=\int_{V}pdV,
\end{eqnarray}
where the integration acts over the volume per unit cylinder length 
$V$, and $m$ is the line density of the central spindle. 
Assuming constant density $\rho$ in the spindle and using
scale radius $R=\sqrt{m/\pi\rho}$, the integration can be worked out:
\begin{equation}
{\cal I}=\frac{1}{2}mR^2,~{\cal T}=\frac{1}{4}m(\frac{dR}{dt})^2, 
~\Pi=\frac{k_B T}{\mu m_{\rm H}}m,
\label{eq:ITP}
\end{equation}
where $\mu$ is the mean molecular weight, and 
$m_{\rm H}$ is the mass of a hydrogen nucleus.
Substituting equation (\ref{eq:ITP}) into equation (\ref{eq:virial}), 
we obtain
\begin{equation}
\frac{d^2R}{dt^2}=-\frac{2G}{R}[m-m_{\rm crit}(T)],
\end{equation}
where the critical line density 
\begin{equation}
m_{\rm crit}(T)=\frac{2k_B T}{\mu m_{\rm H} G}.
\end{equation}
Note that the gas pressure exactly balances the gravity
for the critical line density $m_{\rm crit}(T)$.

The line density of the spindle $m$ is not constant during 
the collapse. 
If the gravity is far larger than pressure, the 
outer part of the spindle is left to the envelope.
In other words, when the line density of the central spindle
$m$ is much larger than the critical value $m_{\rm crit}$, 
$m$ decreases during the collapse of the cylinder.
To mimic this effect, we assume that if $m$ is larger than a 
critical value, $m$ decreases at the same rate as the core radius 
does: 
\begin{equation} 
  \frac{dm}{dt}=(m-m_{\rm th})  \frac{1}{R} \frac{dR}{dt}. 
\end{equation}
Here, we take the threshold line density $m_{\rm th}=1.5m_{\rm crit}$ 
following Nishi (2002), 
who found this treatment reproduces the numerical results
of Nakamura and Umemura (1999) well. 

As for the envelope, we employ an assumption similar to the case 
of spherical cores (section 2.1.1): 
physical quantities of a mass element evolve only when the mass element is
inside the central spindle, and remain constant after the mass
element is left to the envelope.  
Therefore, by calculating only physical quantities of the central spindle, 
we can obtain the whole cloud structure. 

Thermal and chemical processes are treated in the same way 
as in the spherical case. 
The luminosity per unit length is also calculated in a way similar to 
before, by summing the emissivity 
both in the dense spindle and the envelope.
When we compare this with the spherical ones, 
we set the filament mass to be $M_{\rm fil}=10^5\MO$, and
thus length, $l_{\rm fil}=M_{\rm fil}/m_{\rm i}$. 

When the fragmentation timescale of filaments 
$t_{\rm frag}=2.1/(2\pi G\rho)^{1/2}$ (Nagasawa 1987) becomes shorter
than the collapse timescale $t_{\rm dyn}=\rho/(d\rho/dt)$, 
fragmentation is expected to occur. 
We follow the evolution of the filaments up to this epoch.

\subsection{Initial condition}

Collapse and fragmentation of a filamentary primordial gas cloud 
have been investigated by some authors
(e.g., Uehara et al. 1996; Uehara, Inutsuka 2000; Nakamura, Umemura 2002). 
Following the classification of the filaments by Nakamura and Umemura (2002),  
we study three varieties of filaments: 
\begin{itemize}
\item F1: a low-density filament with low ${\rm H_2}$ fraction, 
whose contraction is controlled by ${\rm H_2}$-cooling: 
the initial number density $n_{\rm H}=10^2{\rm cm^{-3}}$, 
temperature $T=400$K, and ${\rm H_2}$ abundance $y({\rm H_2})=10^{-4}$, 

\item F2: a low-density filament with high ${\rm H_2}$ fraction, 
whose contraction is controlled by HD cooling: 
$n_{\rm H}=10^2{\rm cm^{-3}}$, $T=300$K, and $y({\rm H_2})=3\times 10^{-3}$, 

\item F3: a high-density filament:
$n_{\rm H}=10^6{\rm cm^{-3}}$, $T=400$K, and $y({\rm H_2})=10^{-3}$. 
\end{itemize}
We use the initial ionization degree in filaments 
$y({\rm H^+})=y({\rm e^-})=5\times 10^{-5}$ following 
Nakamura and Umemura (2002).
Except ${\rm H_2}$, ${\rm H^+}$, and ${\rm e^-}$, 
we use the same initial abundances of chemical species
as in the case of the spherical cores.

Using the ratio of gravity to pressure gradient $f$, and 
the critical line mass $m_{\rm crit}(T_{\rm init})$ for initial 
temperature $T_{\rm init}$, 
the initial line density $m_{\rm i}$ can be written as 
$m_{\rm init} = f m_{\rm crit}(T_{\rm init})$. 
Here, we assume $f=2$, which is the most probable value for 
filaments formed by fragmentation of a sheet-like cloud 
(Miyama et al. 1987a,b).
For the total mass of $10^5 \MO$, the filament lengths are (F1)32.5 pc, 
(F2)43.4 pc, and (F3) 32.5 pc.

\subsection{Results} 
\subsubsection{Thermal and Chemical Evolution}
\begin{figure}
  \begin{center}
    \FigureFile(80mm,80mm){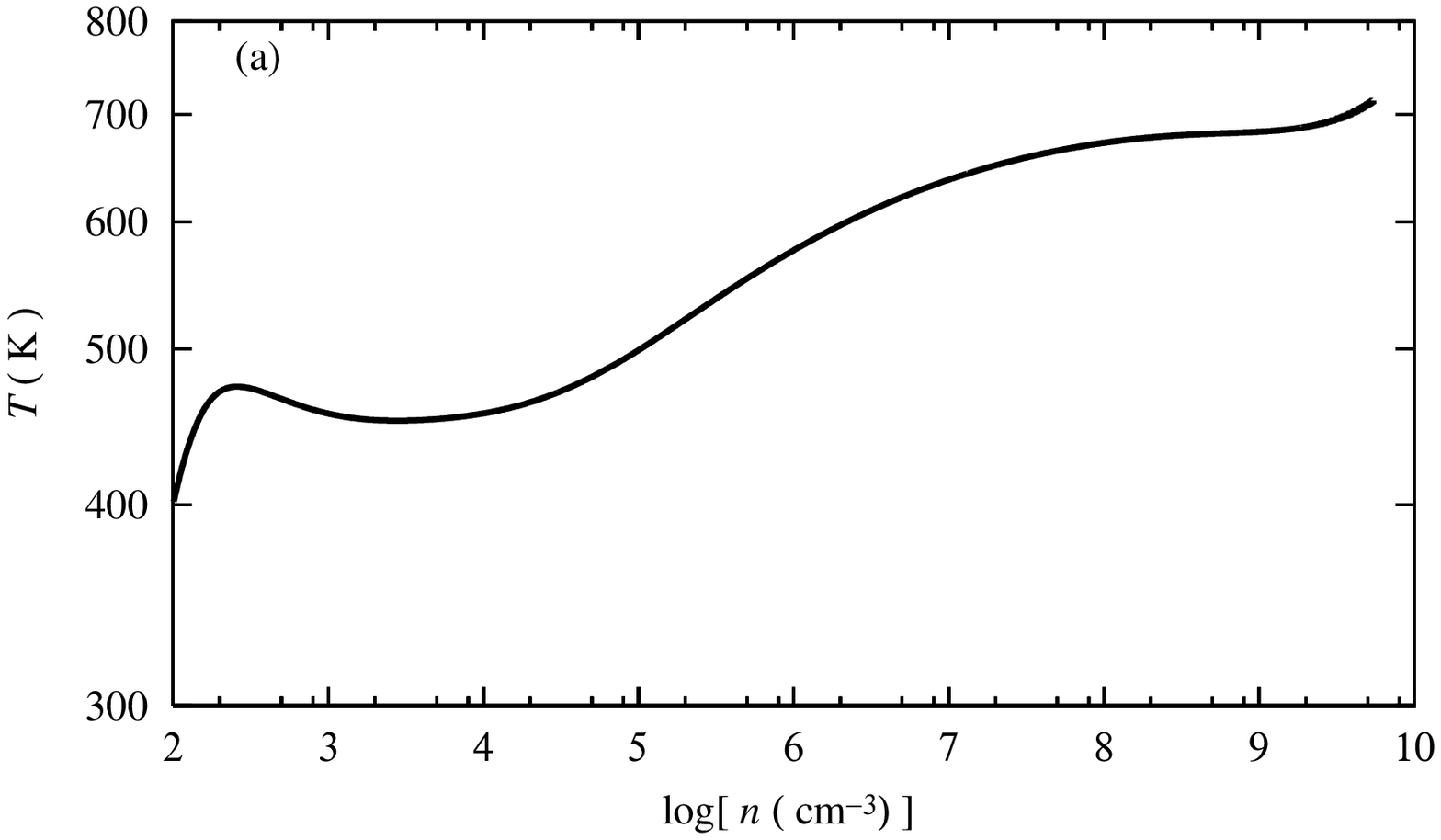}
    \FigureFile(80mm,80mm){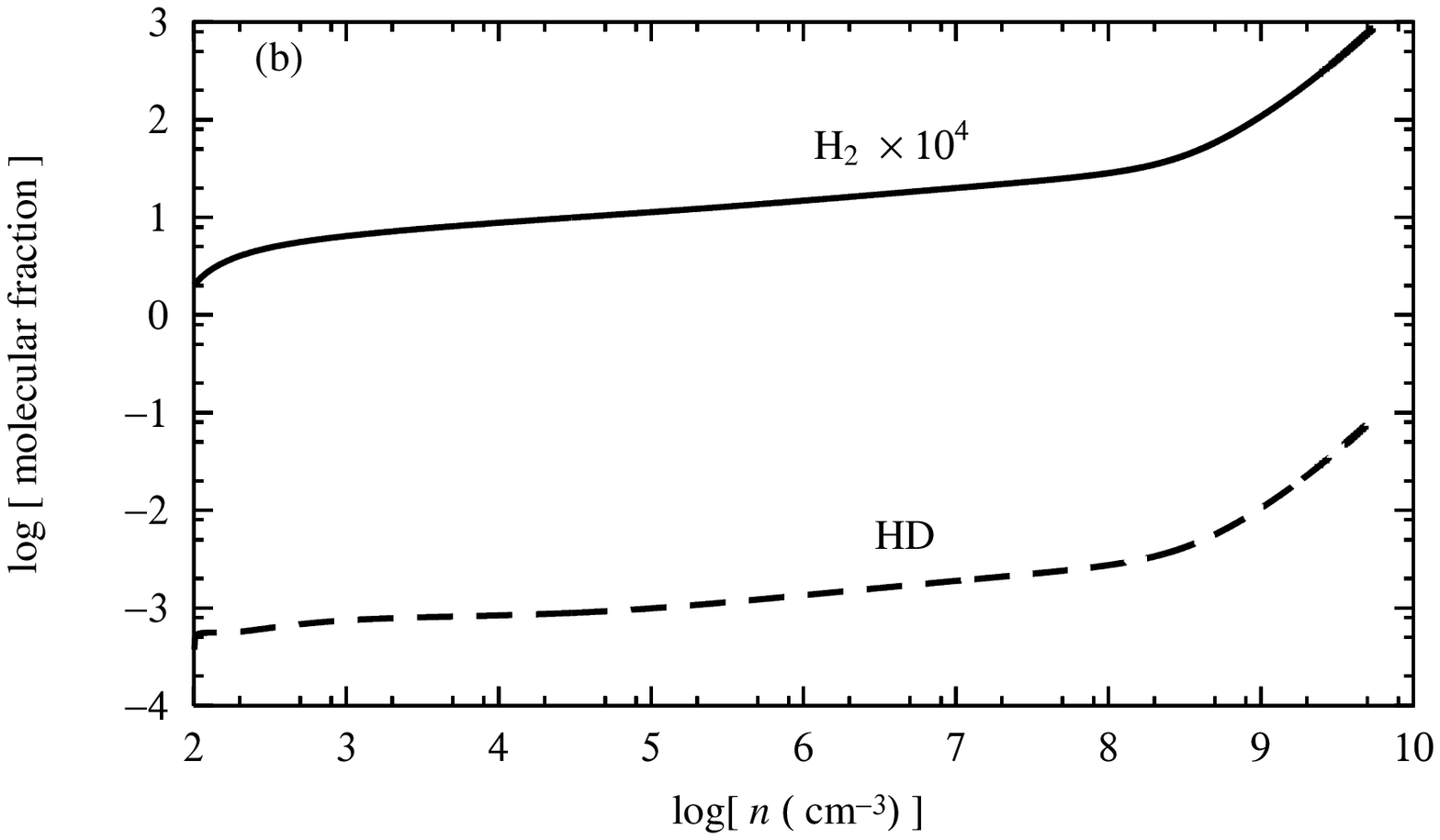}
    \FigureFile(80mm,80mm){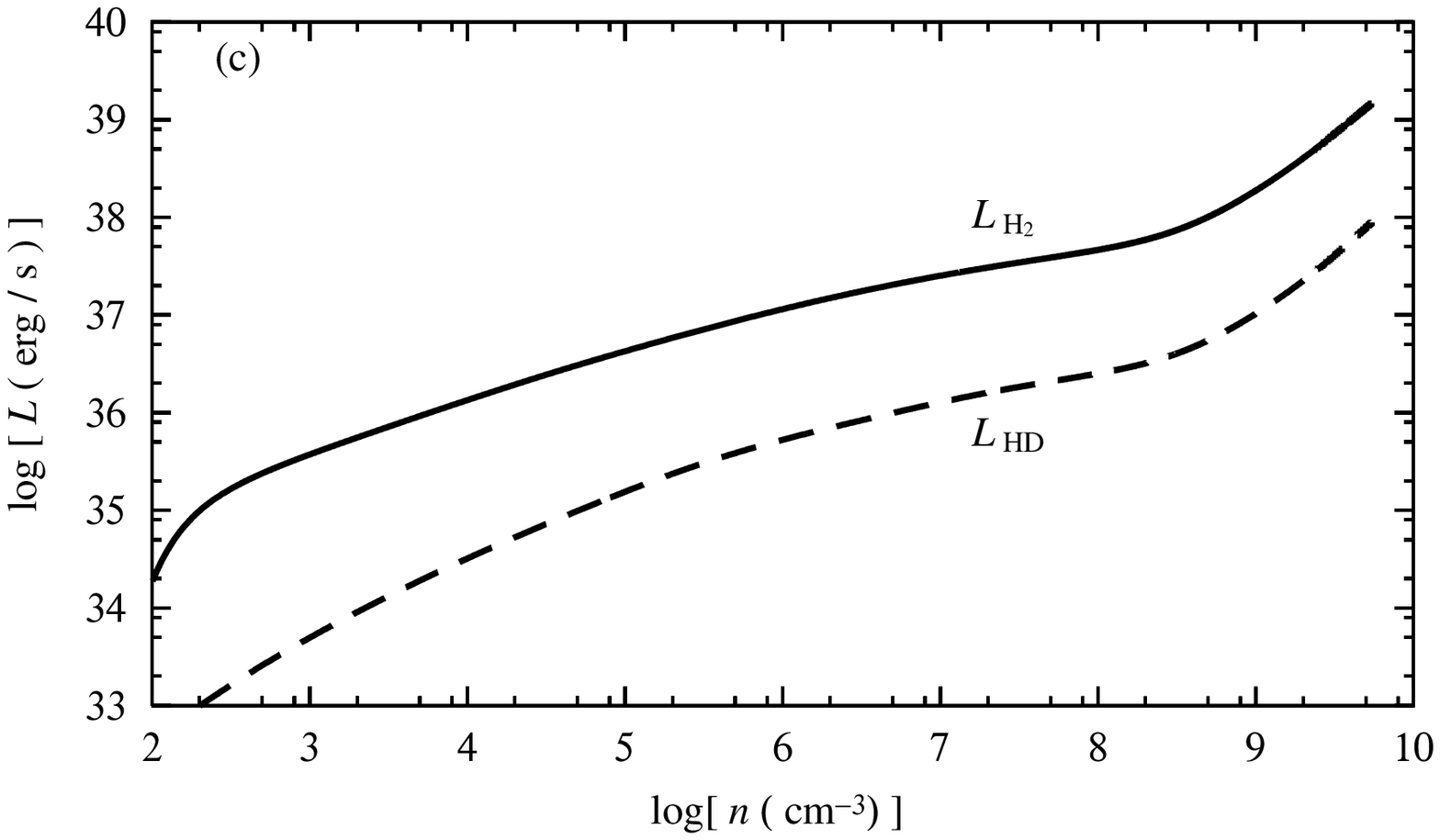}
  \end{center}
\caption{Evolution of the filamentary cloud of model F1, where
the contraction is induced by the H$_2$ cooling.
Shown are 
(a) the temperature,  
(b) the H (solid) and D (dashed) molecular fractions and 
(c) the cooling rate by ${\rm H_2}$ line (solid) and 
HD line (dashed) 
at the center. }
\label{fig:TH2energy_cylA}
\end{figure}

\begin{figure}
  \begin{center}
    \FigureFile(80mm,80mm){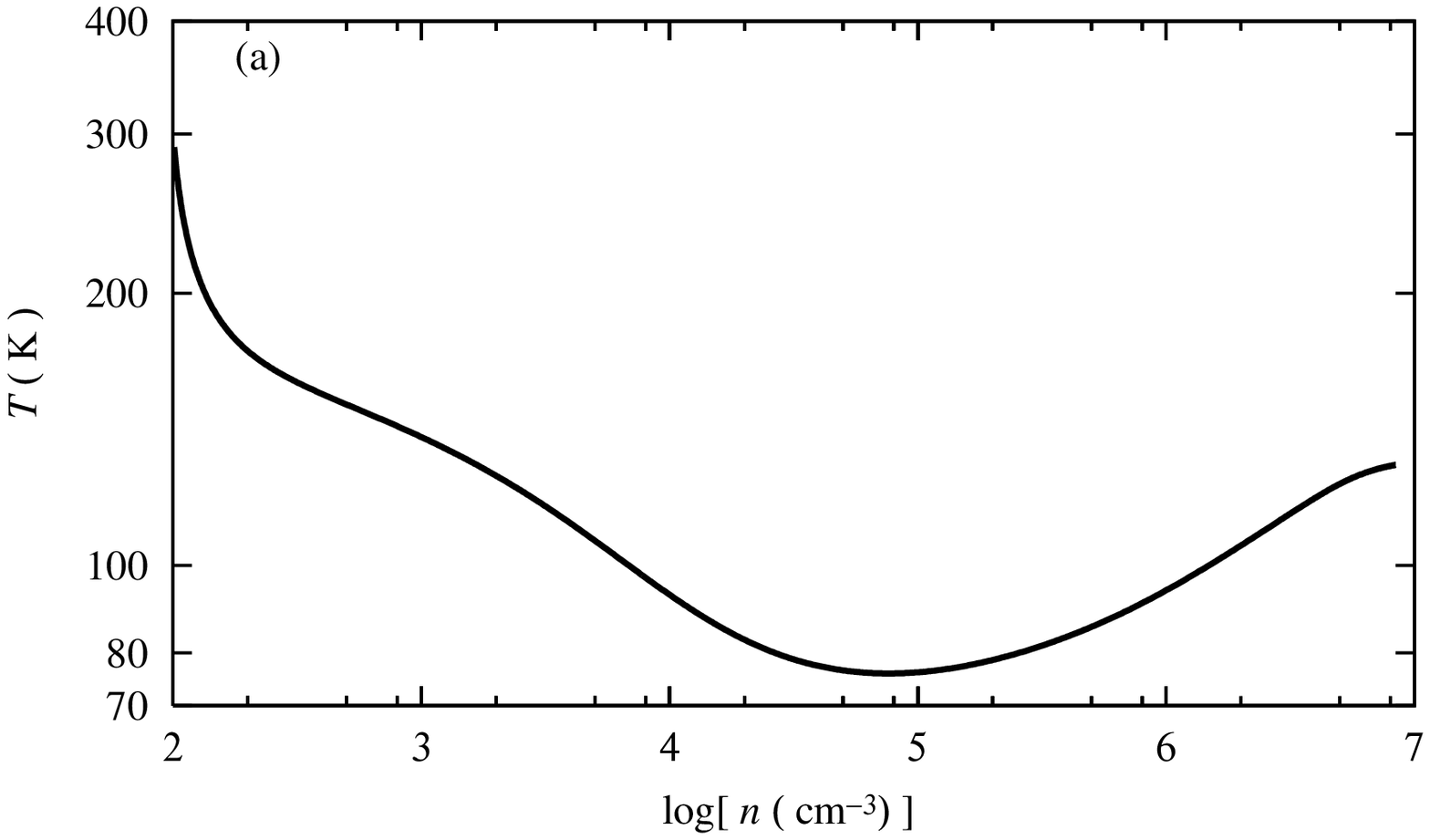}
    \FigureFile(80mm,80mm){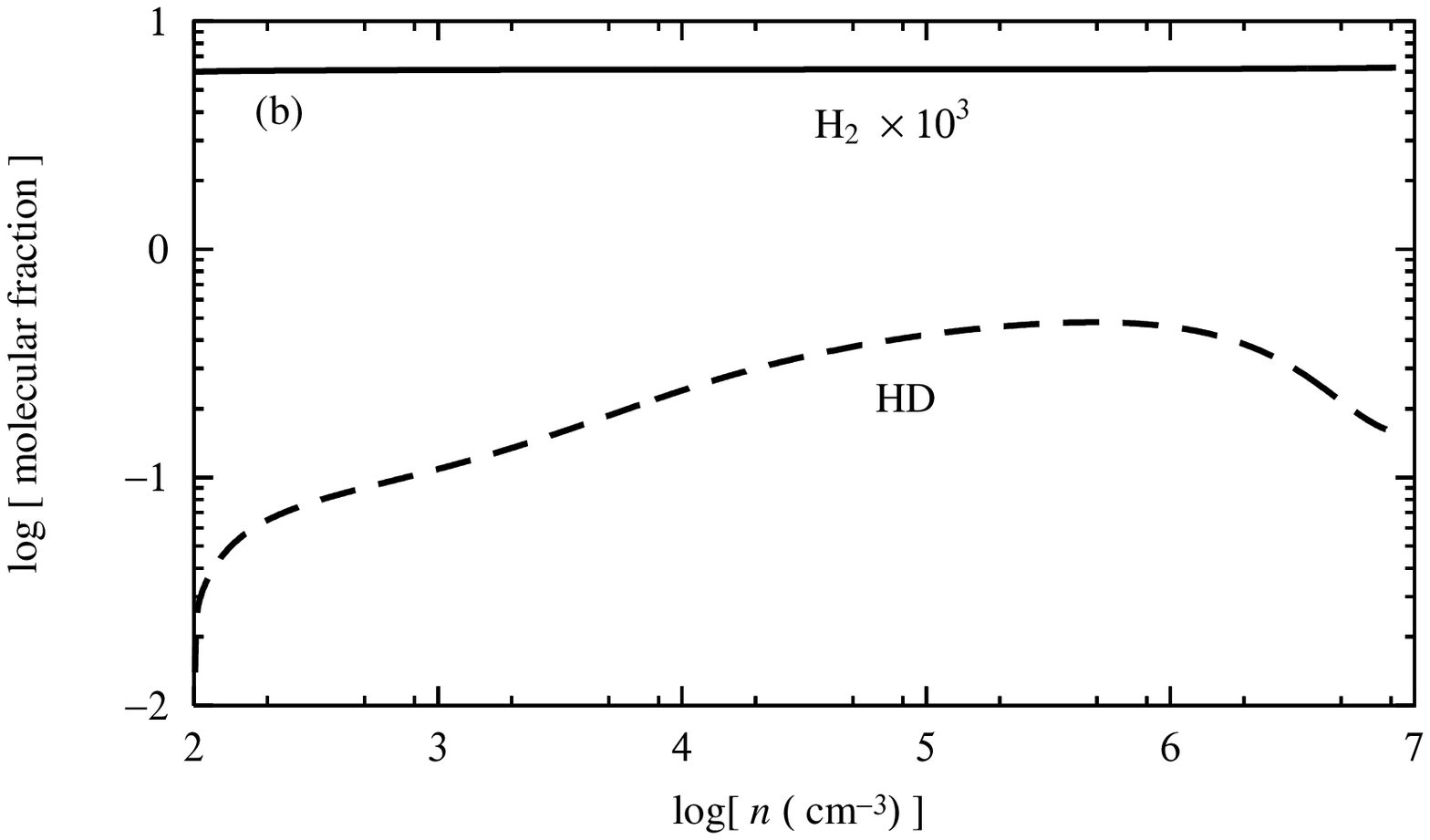}
    \FigureFile(80mm,80mm){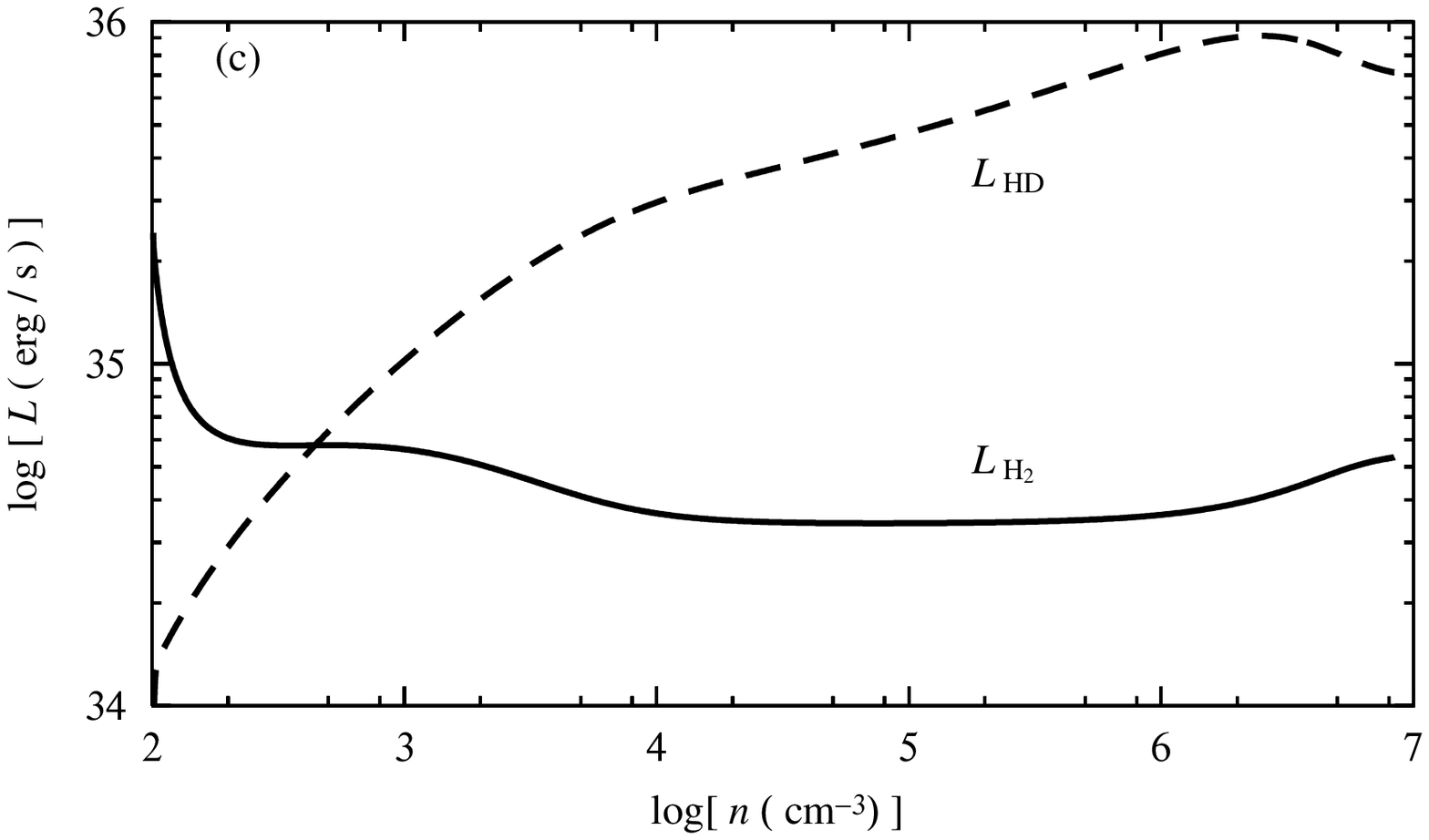}
  \end{center}
\caption{Evolution of the filamentary cloud of model F2, where
the contraction is induced by the HD cooling.
Meaning of the panels is the same as figure \ref{fig:TH2energy_cylA}.}
\label{fig:TH2energy_cylB}
\end{figure}

\begin{figure}
  \begin{center}
    \FigureFile(80mm,80mm){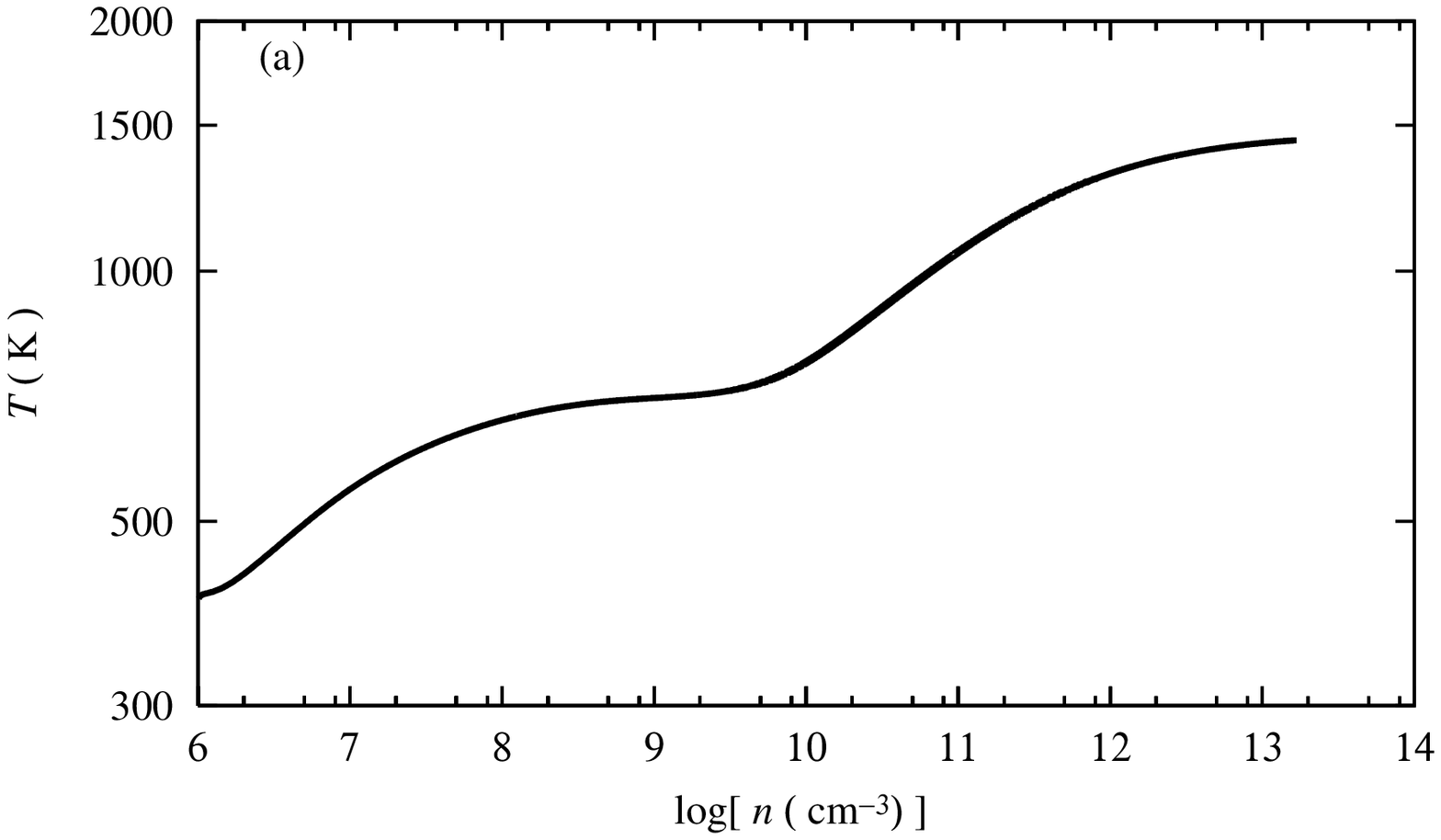}
    \FigureFile(80mm,80mm){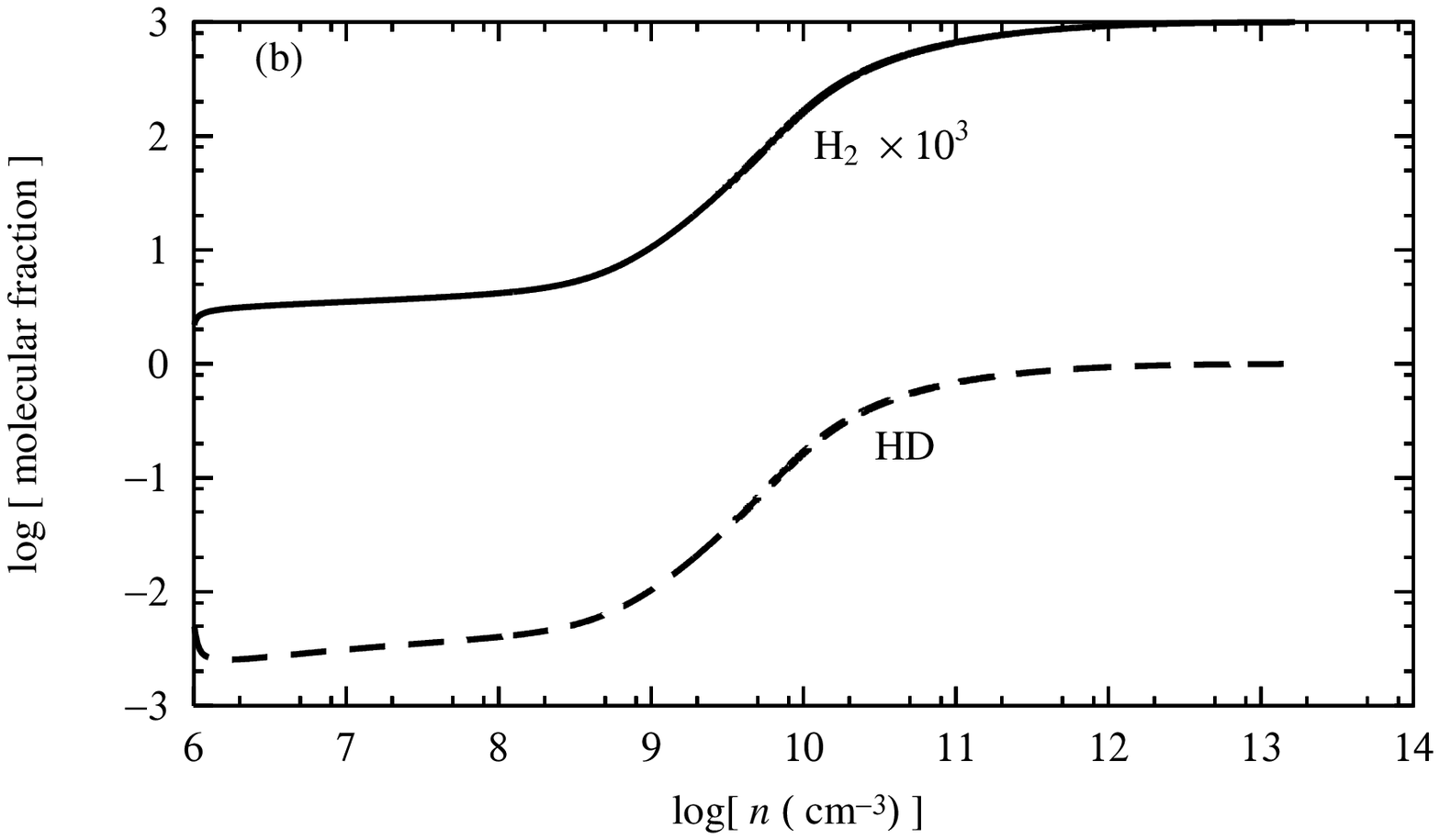}
    \FigureFile(80mm,80mm){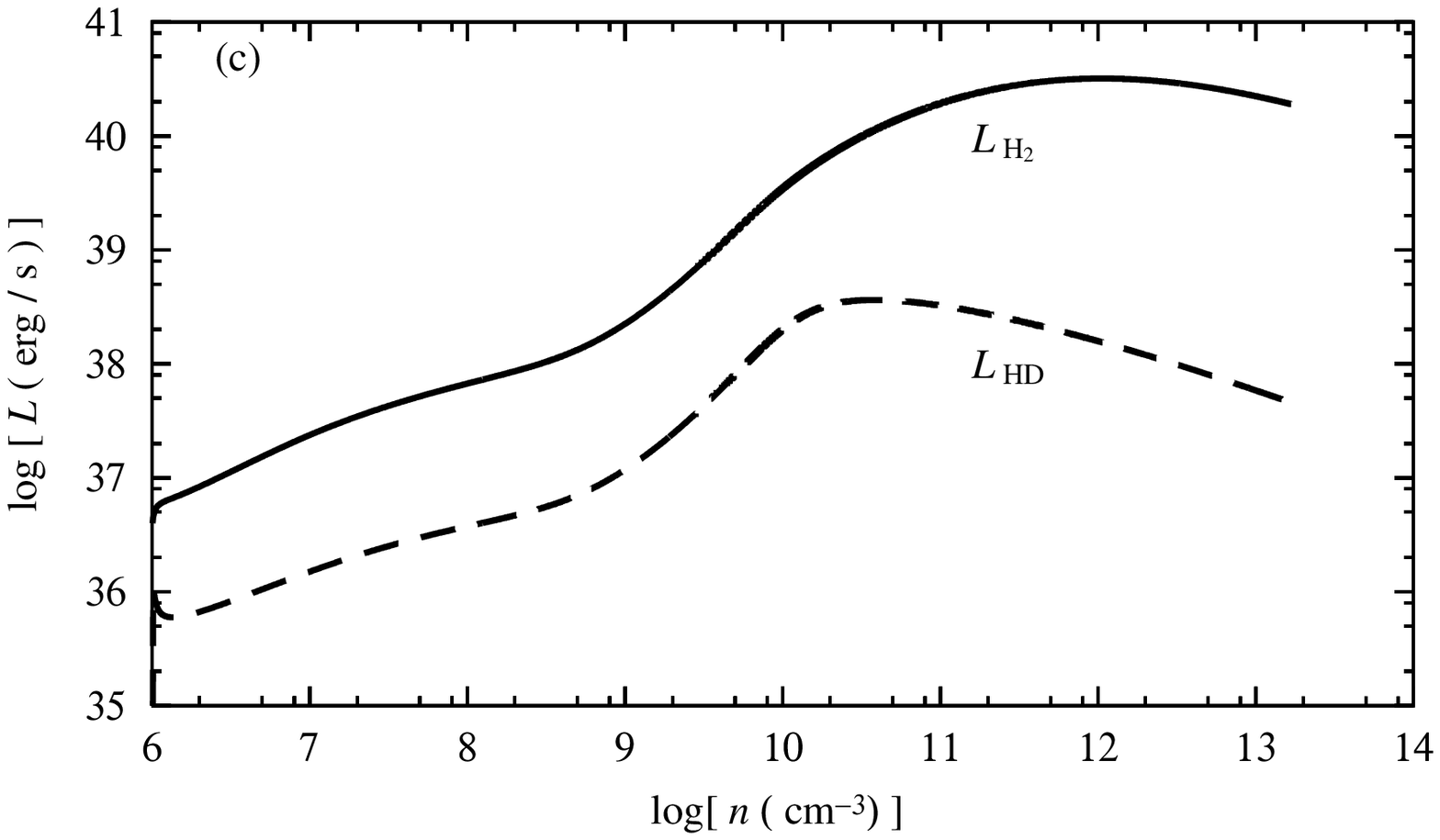}
  \end{center}
\caption{Evolution of the filamentary cloud of model F3, the case of 
a high-density filament.
Meaning of the panels is the same as figure \ref{fig:TH2energy_cylA}.}
\label{fig:TH2energy_cylC}
\end{figure}

Figures \ref{fig:TH2energy_cylA}-\ref{fig:TH2energy_cylC} 
show evolution of (a) temperature; 
(b) abundances of ${\rm H_2}$ (solid line) and HD (dashed line); 
(c) cooling rates by ${\rm H_2}$ $L_{\rm H_2}$ (solid line) and 
by HD $L_{\rm HD}$ (dashed line). 
We do not present the evolution of LiH abundance here, 
since LiH formation proceeds in a fashion similar to  
the case of the spherical cores: 
only a tiny fraction of Li becomes molecular form 
without affecting thermal evolution.

The overview of evolution in each case is as follows:

\begin{itemize}
\item F1: the low-density filament with low ${\rm H_2}$ fraction

The evolution resembles that of the spherical cores 
with low ${\rm H_2}$ fraction (e.g., model S1 in section 2). 
HD fraction remains low (Figure \ref{fig:TH2energy_cylA}b).  
Throughout the contraction, dominant cooling mechanism is H$_2$ line cooling, 
which nearly balances the compressional heating 
(Figure \ref{fig:TH2energy_cylA}c). 
In densities higher than the critical density for collisional 
de-exitation of H$_2$ rotational level ($\sim 10^{4}$K), 
temperature rises slowly from 400K to 700K 
(Figure \ref{fig:TH2energy_cylA}a,c). 
With the onset of rapid H$_2$ formation via the three-body reactions
and resultant heat injection, the temperature rises suddenly and 
the pressure becomes stronger than 
the gravity (Figure \ref{fig:TH2energy_cylA}c).
At this moment ($\sim 10^{10}{\rm cm^{-3}}$) the fragmentation 
timescale $t_{\rm frag}$ becomes shorter than the collapse 
timescale $t_{\rm dyn}$ and then the cylinder fragments. 

\item F2: the low-density filament with high ${\rm H_2}$ fraction

As in the case of the star-forming cores 
with high ${\rm H_2}$ fraction (model S5, 6 in section 2), 
HD is formed abundantly and becomes an important coolant 
(Figure \ref{fig:TH2energy_cylB}a,c). 
As a result, the temperature goes down to $<100$K.  
When the filament contracts $n \gtrsim 10^5{\rm cm^{-3}}$, 
which is the critical density for HD collisional de-excitation, 
the temperature begins rising as described in section 2.3.1. 
Owing to this temperature increase, HD begins to dissociate.
This results in further temperature increase and the fragmentation 
occurs at $<10^7{\rm cm^{-3}}$.

\item F3: the high-density filament
 
The initial density being already higher than the LTE critical 
density for ${\rm H_2}$ and HD ($\sim 10^4{\rm cm^{-3}}$ 
for H$_2$ and $\sim 10^5{\rm cm^{-3}}$ for HD), the temperature increases 
with increasing density.

This filament begin to collapse at rather high density
from a highly gravitationally unstable ($f=2$) state.
Even in a density range as high as $\sim 10^8{\rm cm^{-3}}$, the gravity 
dominates over the pressure, and the filament contracts faster 
than in other models.
Then, in spite of H$_2$ formation heating around $n\gtrsim 10^8{\rm cm^{-3}}$,
this filament contracts without fragmenting until $\sim 10^{13}{\rm cm^{-3}}$
where it becomes opaque to the ${\rm H_2}$ lines.

\end{itemize}

\subsubsection{Luminosities from the Filaments}

\begin{figure}
  \begin{center}
    \FigureFile(90mm,80mm){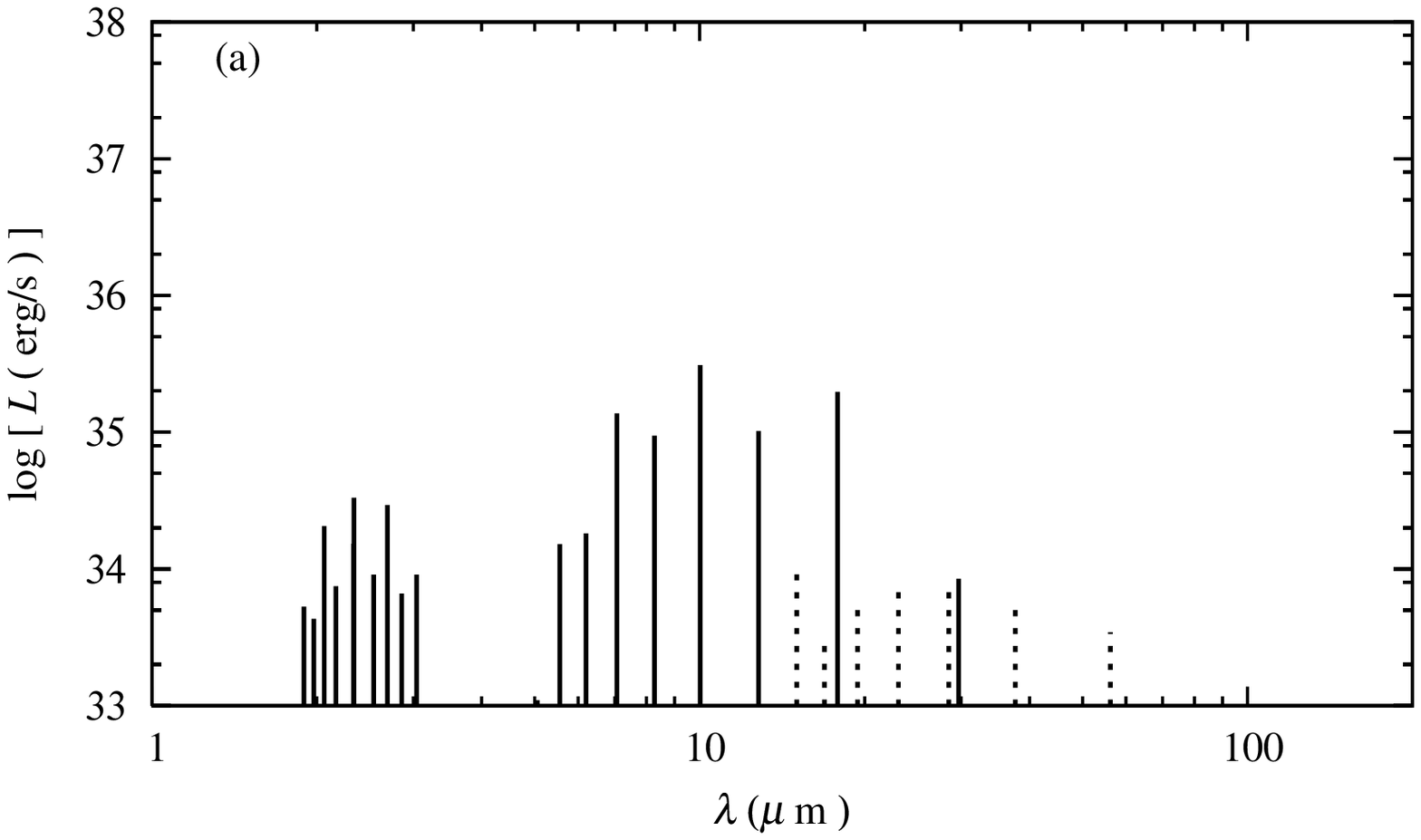}
    \FigureFile(90mm,80mm){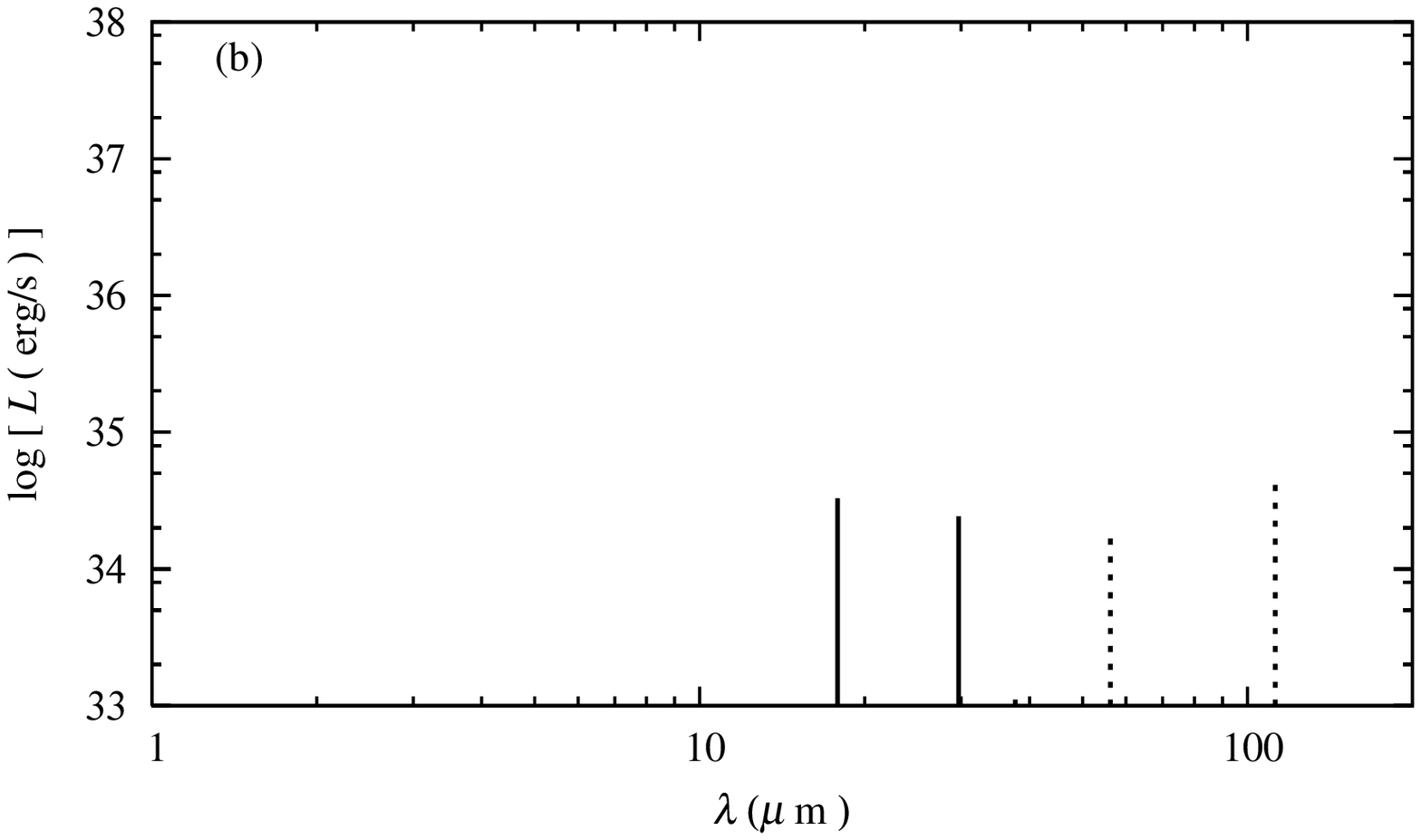}
    \FigureFile(90mm,80mm){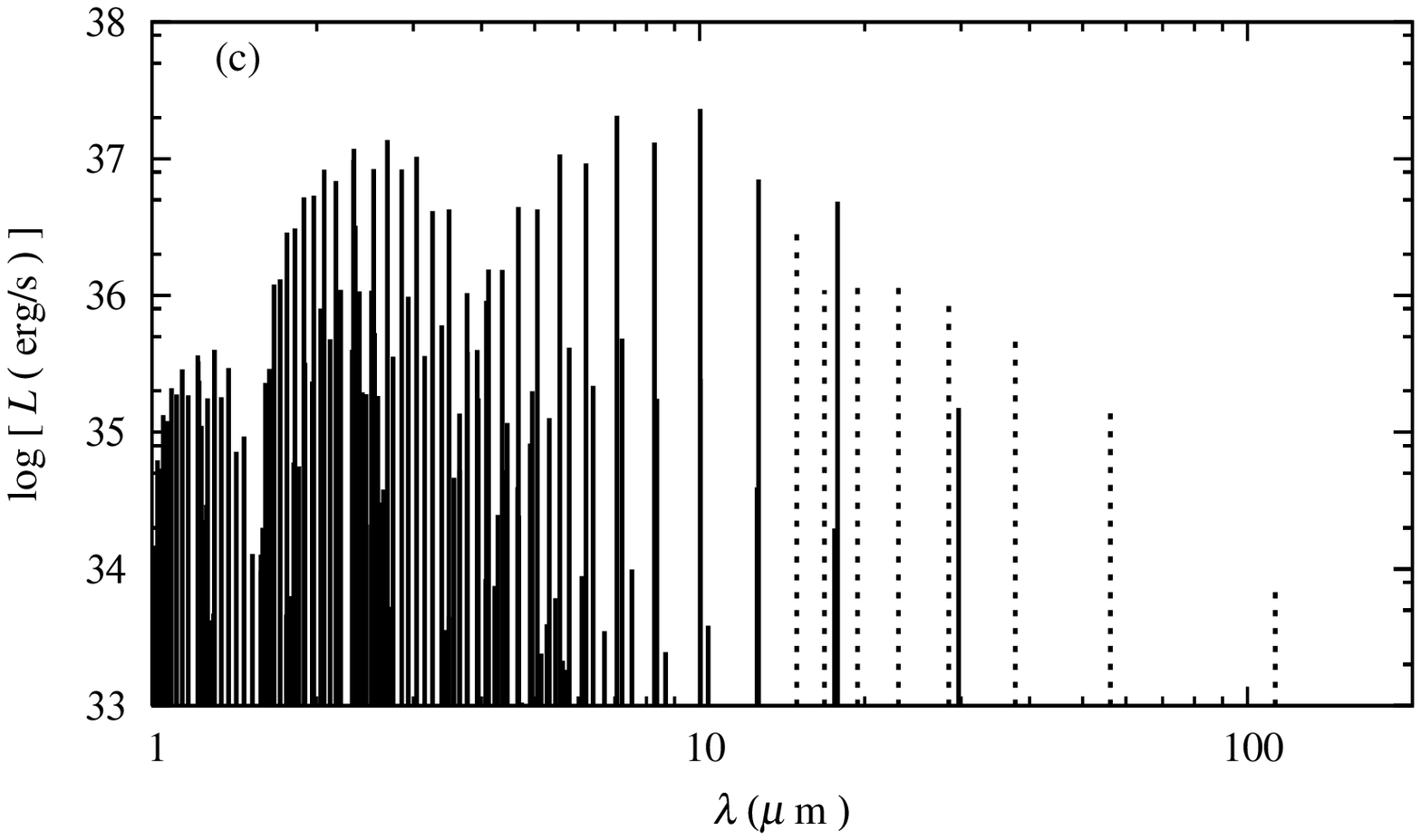}
    \FigureFile(90mm,80mm){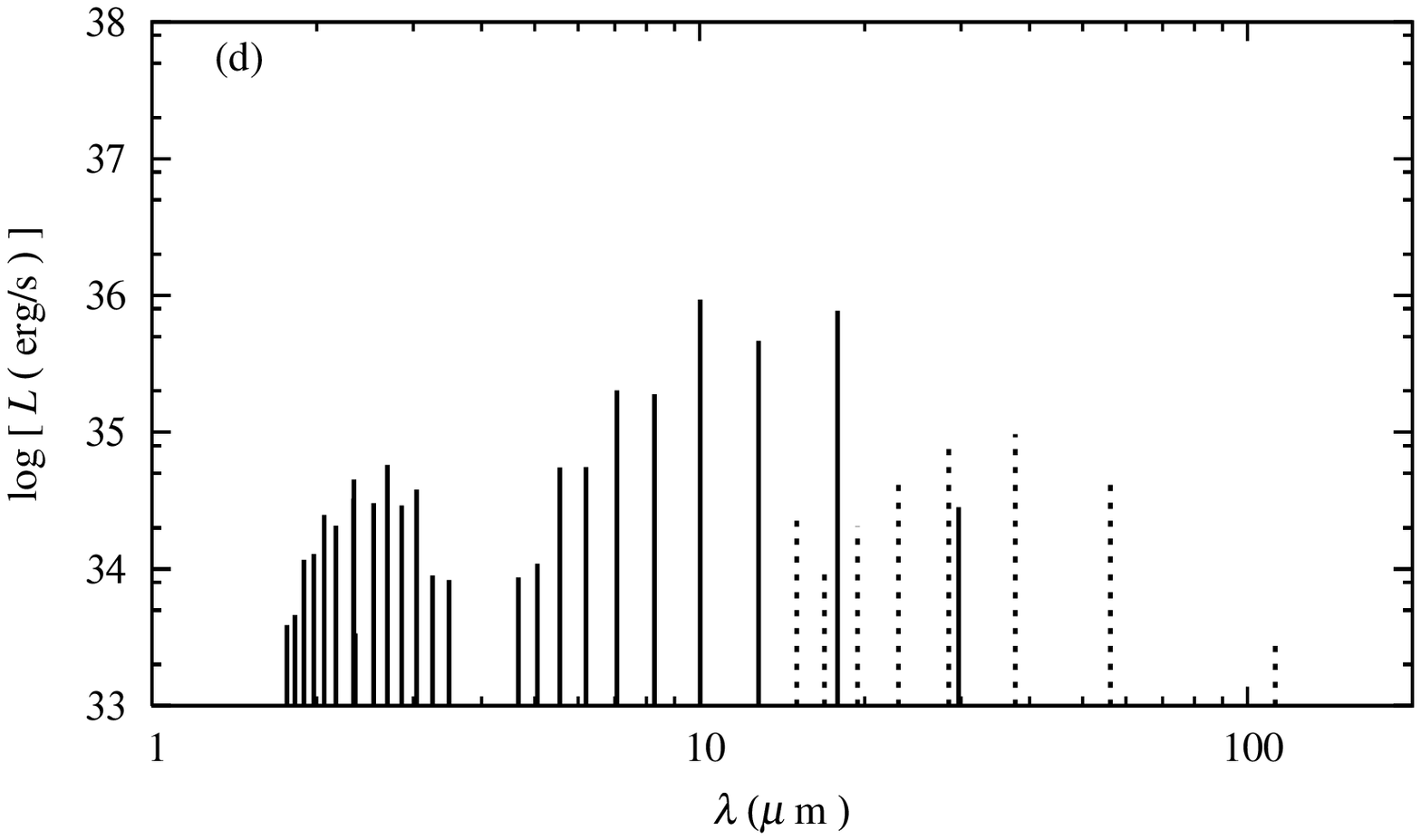}
  \end{center}
\caption{Luminosities of individual ${\rm H_2}$ lines (solid lines) and 
HD lines (dotted lines) averaged over the contraction time. 
Shown are for models (a) F1, (b) F2, and  (c) F3. 
The mass is normalized to $10^5\MO$ for all cases.
Also, in panel (d), the case of a spherical core in the collapse phase 
with the same initial condition as model F3 
is shown.}
\label{fig:L(time)cyl}
\end{figure}

In Figure \ref{fig:L(time)cyl}, time-averaged luminosities 
of ${\rm H_2}$ and HD lines from the filamentary clouds
are shown.
Here, time averaging is over the total contraction time before 
fragmentation. 
Total mass of the filaments is normalized to $10^5 \MO$. 

In model F3 (Figure \ref{fig:L(time)cyl}c),
the filament evolves in the region of high temperature 
(400K$\lesssim T \lesssim $1500K) and high density
($10^6{\rm cm^{-3}}\lesssim n \lesssim 10^{13}{\rm cm^{-3}}$) 
where ${\rm H_2}$ cooling is efficient.
Accordingly, strong radiation is emitted with 
some H$_2$ lines exceeding $10^{37} {\rm erg~s^{-1}}$ 
and HD lines exceeding $10^{36}{\rm erg~s^{-1}}$. 
Luminosities are weaker for model F1 where all the 
line luminosities are less than 
$10^{36}{\rm erg~s^{-1}}$ (figure \ref{fig:L(time)cyl}a)
because of the lower temperature (400-700K) in model F1 
throughout the contraction.
With temperature even lower ($T<$200K),  in model F2 
only a few lower-lying rotational lines of ${\rm H_2}$ and HD 
are stronger than $10^{34} {\rm erg~s^{-1}}$.
Since HD is more effective as a coolant than H$_2$,  
most of the ${\rm H_2}$ lines are very faint ($<10^{32} {\rm erg~s^{-1}}$). 

For the purpose of seeing the effect of cloud geometry,  
we show in Figure \ref{fig:L(time)cyl}(d) the mean luminosities 
of spherical star-forming cores of total mass $10^5 \MO $ calculated
with the same initial condition as for model (F3).
We can see that line luminosities of model (F3) are stronger than 
those of star-forming cores with the same mass
by about an order of magnitude.
The reason is as follows: 
In the case of a spherical core, the fully molecular region, where
H$_2$ emission is strongest owing to heating by H$_2$ formation,
accounts for at most $\sim 1 \MO$ at the center (Omukai, Nishi 1998). 
Most of the matter is left to the envelope before this epoch.
On the other hand, in the cylindrical case, 
the fraction of envelope mass is less than 
that in the case of the run-away collapse phase. 
Although the radius of the fully molecular region
at the center of the cylinder is almost the same as in the spherical case, 
its length in the axis direction is longer.
Therefore, the mass of the fully molecular spindle is larger
and emission from it is stronger.

\section{Discussion}
\subsection{Analytical Expression for the Luminosity}

In this section, we derive an analytical expression for the luminosity from 
the collapsing clumps, which reproduces well our numerical results.
First, let us consider the luminosity from a spherical core 
in the collapse phase. 
During the run-away collapse phase, 
length scale of the central region is about a Jeans length 
$ \lambda_{\rm J}$ and mass of the region $M_{\rm cen}$ is about a 
Jeans mass $M_{\rm J}\equiv \rho \lambda_{\rm J}^3$.
Using cooling rate (per unit volume) in the central region  
$\Lambda_{\rm cen}$, the luminosity from the central region 
is given by 
\begin{equation}
L_{\rm cen} \simeq M_{\rm cen} \Lambda_{\rm cen}/ \rho_{\rm cen}. 
\label{eq:L_cen0}
\end{equation}
Here we assume an energy balance between heating and cooling, 
which is valid in a good accuracy during the run-away collapse phase:
\begin{equation}
\Lambda_{\rm cen} \simeq \Gamma_{\rm cen}.
\label{eq:LeqG}
\end{equation}
Also, we consider only the compression as a source of heating, then 
\begin{equation}
\frac{\Gamma_{\rm cen}}{\rho_{\rm cen}} 
\simeq 
-P_{\rm cen} \frac{d}{dt}\left( \frac{1}{\rho_{\rm cen}} \right)
\simeq \frac{k_{\rm B}T}{\mu m_{\rm H}} \frac{1}{t_{\rm ff}},
\label{eq:heat}
\end{equation}
where we have assumed that the timescale of contraction is approximated 
by the free-fall time. 
Substituting equations (\ref{eq:LeqG}) (\ref{eq:heat}) into (\ref{eq:L_cen0}),
the luminosity from the central region is given by
\begin{equation}
L_{\rm cen} \simeq \sqrt{\frac{32}{3\pi}}\frac{c_{\rm s}^5}{G}
\simeq 46L_{\odot}\left(\frac{T}{1000K}\right)^{\frac{5}{2}} {\rm erg~s^{-1}}.
\label{eq:L_cen}
\end{equation}
Note that $L_{\rm cen}$ is a function only of temperature:
it does not depend, for example, on mass or density of the central region. 
In other words, if the temperature is constant throughout the evolution, 
the luminosity from the central region remains constant.

To calculate the total luminosity, we need to add the contribution 
from the envelope. 
In the collapse phase, the Jeans mass decreases with increasing density. 
A mass shell initially contained within the central region at some time
leaves the central evolution and becomes part of the envelope 
when the mass inside the shell equals the central Jeans mass.
We assume that physical quantities in the envelope are frozen:
physical quantities on a mass shell are kept constant 
after the shell leaves the central region 
until the end of the collapse phase.
Then, the cooling rate of a mass shell in the envelope is given by that of
the central region at the moment when the shell is detached from there.
Because of homogeneity inside the central region, 
the cooling rate in this region is
\begin{eqnarray}
\left( \frac{\Lambda}{\rho}\right)_{\rm cen}= \frac{L_{\rm cen}}{M_{\rm cen}}
\end{eqnarray}
where $L_{\rm cen}$ and $M_{\rm cen}$ is the luminosity and mass of the
central region.
Since a shell that contains mass $M$ inside is detached from the central 
region when $M_{\rm cen}=M$,
the cooling rate of the mass shell in the envelope $\Lambda(M)$ is given by
\begin{eqnarray}
\frac{\Lambda}{\rho}(M) = \frac{L_{\rm cen}}{M}.
\end{eqnarray}
Here we use the constancy of luminosity of the central region
i.e., $L_{\rm cen}(M)=L_{\rm cen}$ (see equation \ref{eq:L_cen}),
considering a star-forming core collapsing isothermally.
Summing over all the mass shells in the envelope, the luminosity of 
the envelope is 
\begin{eqnarray}
L_{\rm env}&=&\int_{M_{\rm cen}}^{M_{\rm tot}}\frac{\Lambda}{\rho}(M) dM \\
&=&L_{\rm cen} {\rm ln}\left(\frac{M_{\rm tot}}{M_{\rm cen}}\right),
\end{eqnarray}
where $M_{\rm tot}$ is the total mass of the star-forming core.
Then, the total luminosity of the star-forming core is
\begin{eqnarray}
L_{\rm tot}&=&L_{\rm cen}+L_{\rm env}\\
&=&L_{\rm cen}
\left[1+{\rm ln}{\left(\frac{M_{\rm tot}}{M_{\rm cen}}\right)}\right].
\end{eqnarray}
The envelope contribution enhances the total luminosity 
only logarithmically:
with increasing central density and decreasing $M_{\rm cen}$, 
total luminosity increases only very slowly.
In the following, we do not discriminate between $L_{\rm cen}$ and 
$L_{\rm tot}$.
Recall that, in deriving the expression above, 
we have not specified a mechanism for cooling. 
In the case of primordial gas, which we consider here, 
cooling is via H$_2$ and HD line cooling. 
If many lines are equally important for cooling, the total 
luminosity is divided into them. 
However, in many situations, one or two lines become dominant. 
In such cases, the luminosities of these lines are given 
approximately by the total luminosity $L_{\rm tot}$. 
From equation (\ref{eq:L_cen}), we also see that 
when HD becomes more important than H$_2$ in cooling, 
low temperature ($\sim$ 100K) causes a corresponding 
drop in luminosity.

\begin{figure}
  \begin{center}
    \FigureFile(80mm,80mm){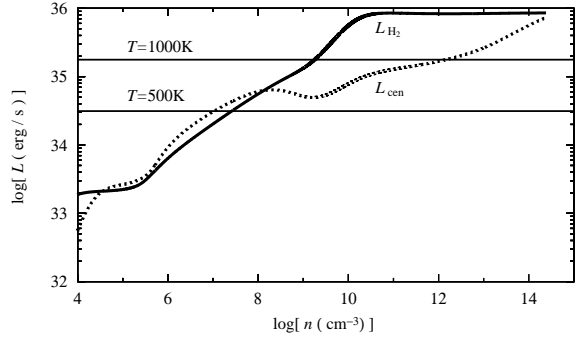}
  \end{center}
\caption{The luminosity in the ${\rm H_2}$ lines (solid) 
for a spherical core (the fiducial model S1) 
in the collapse phase against the number density.
Dotted and thin lines show the values by the analytical expression 
(equation \ref{eq:L_cen}) 
for the temperature evolution in model S1 and for constant temperatures of
$T=500$K and $1000$K.
}
\label{fig:energy_sp}
\end{figure}

Here we compare the analytical expression (\ref{eq:L_cen}) with the
numerical results in section 2.
In Figure \ref{fig:energy_sp}, we show the H$_2$-line luminosity 
$L_{\rm H_2}$ for model S1 and the analytical values (\ref{eq:L_cen}) 
for the model S1 temperature evolution.
Also shown are the analytical values for constant temperatures 
$T=500$K and $1000$K.
For $n<10^8{\rm cm^{-3}}$, the numerical value $L_{\rm H_2}$ matches well 
the analytical value $L_{\rm cen}$.
On the other hand, the numerical result deviates from
the analytical value for higher densities ($n>10^8 {\rm cm^{-3}}$)
because the assumption that the main heating
source is gravitational compression, which is postulated in
deriving equation  (\ref{eq:L_cen}), is violated:
for these high densities,  dominant heating process is the chemical
energy released by H$_2$ formation.
In particular, the strongest emission comes from the region of
$n=10^9-10^{10}{\rm cm^{-3}}$, where H$_2$ formation is most active.
For higher densities ($n>10^{10}{\rm cm^{-3}}$),
since majority of hydrogen has already become the molecular form,
the chemical heating decreases and has little contribution to the entire
luminosity.
Then, when the central density is larger than $10^{10}{\rm cm^{-3}}$, 
the total luminosity is dominated by the contribution from the envelope
of $n=10^9-10^{10}{\rm cm^{-3}}$.
The constancy of luminosity for $n>10^{10}{\rm cm^{-3}}$ is the result of
the dominance of the emission from the envelope, which hardly evolve during
the collapse of the center.

Next, we discuss the case of filamentary clouds.
The size of the central spindle is again about a Jeans 
length $\lambda_{\rm J}$ and the line density of the spindle is 
$\sim \rho_{\rm cen} \lambda_{\rm J}^2$.
The contribution of luminosity from the central spindle is then
\begin{eqnarray}
L_{\rm sp} \simeq  \rho_{\rm cen} \lambda_{\rm J}^2 l
(\Lambda_{\rm cen}/ \rho_{\rm cen}),
\label{eq:L_fil} 
\end{eqnarray}
where $l$ is the filament length. 
As in the case of spherical cores, the thermal balance 
(equation \ref{eq:LeqG}) is maintained during the collapse.
If heating is due only to compression, 
using equations (\ref{eq:LeqG}), (\ref{eq:heat}) and (\ref{eq:L_fil}), 
the luminosity of the spindle is given by 
\begin{equation}
L_{\rm sp} \simeq \rho_{\rm cen} \lambda_{\rm J, cen}^2 l 
\frac{k_{\rm B}T}{\mu m_{\rm H}} \frac{1}{t_{\rm ff}}
\simeq 4.2 \times 10^3 L_{\odot}\left(\frac{T}{500K}\right)^2
\left(\frac{n}{10^6 {\rm cm^{-3}}}\right)^{\frac{1}{2}}
\left(\frac{l}{100 {\rm pc}}\right).
\label{eq:L_fil2}
\end{equation}
Unlike spherical cores, the luminosity of
a filamentary cloud rises with increasing density 
even in the isothermal case.
So far, we have considered only the contribution 
from the central spindle. 
In the isothermal case,
since the line density of the central spindle, 
$\rho_{\rm cen} \lambda_{\rm J}^2$, is constant,
all the matter in the filament contracts uniformly without being left 
to the envelope. 
The central contribution $L_{\rm sp}$ becomes equal to the 
total luminosity. 
Indeed, the filaments remain at approximately constant temperature in many 
cases in section 3. 
Then, we assume in the following that $L_{\rm tot}\simeq L_{\rm fil}$.

\begin{figure}
  \begin{center}
    \FigureFile(80mm,80mm){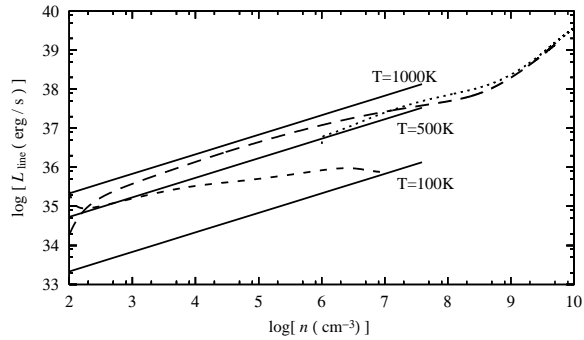}
  \end{center}
\caption{Total luminosity for filaments of model F1 (long-dashed; 
a low-density filament with efficient H$_2$ cooling), 
F2 (short-dashed; a low-density filament with efficient HD cooling), 
F3 (dotted; a high-density filament). 
Also shown by solid lines are the analytical values 
(equation \ref{eq:L_fil}) for 
$T=100{\rm K}, 500{\rm K}$, and 1000K.}
\label{fig:energy_fil}
\end{figure}

Next, we compare the analytical expression (\ref{eq:L_fil2}) 
with the numerical results.
We consider the range $n<10^8{\rm cm^{-3}}$, where H$_2$ formation 
heating is not important. 
In figure \ref{fig:energy_fil}, luminosity evolution  
of models F1-3 is shown, along with 
analytical values shown by equation (\ref{eq:L_fil2}) 
for $T=100{\rm K}$, $500{\rm K}$, and $1000{\rm K}$.
The temperatures in models F1 and F3 are about 500K and remain roughly 
constant, while in model F2 the temperature is about 100K, but with 
significant variation.
For model F1, the analytical value matches the numerical one 
by a factor $<3$.
Agreement between them is even better for model F3.
In the case of model F2, the numerical luminosity deviates from 
the analytical one for 100K because of a large variation in temperature.  
When we calculate the analytical value using 
the temperature evolution of the numerical result,
our numerical result agrees with the analytical one within a factor of a few.
In summary, the analytical estimates reproduce our numerical calculation
within a factor of a few. 

\subsection{Expected Luminosity from Pop III Star-Forming Galaxies}

In this section, we discuss the condition of the Pop III star-formation 
rate (SFR) $\psi_{\rm III}$ and formation redshift $z$ for enabling 
the detection of molecular emission by
next generation facilities, in particular, 
JWST (optical--mid-infrared wavelengths) and 
SPICA (mid--far-infrared wavelengths).
Hereafter, we consider Pop III star-forming galaxies where 
star formation proceeds through 
low-density filaments or through high-density filaments. 
	
\begin{figure}
  \begin{center}
    \FigureFile(80mm,80mm){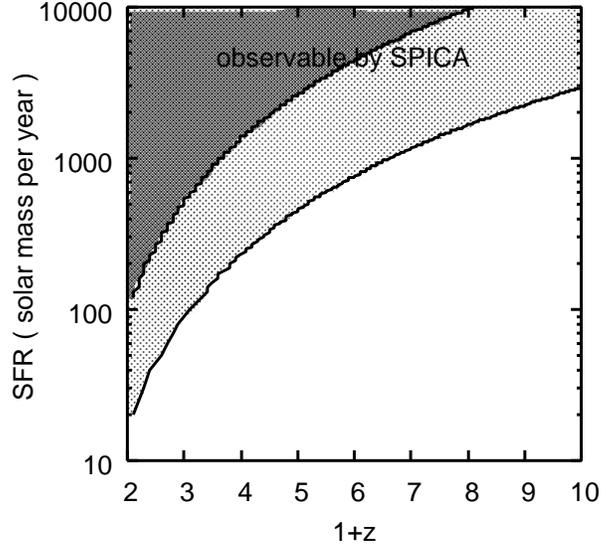}
  \end{center}
\caption{The redshift-star formation rate (SFR) domain 
where the strongest H$_2$ line flux 
$F=L_{\rm cl}/4\pi D_{1+z}^2$ 
exceeds the detection limit of SPICA, $\sim 10^{-22} {\rm Wm^{-2}}$.
The dark (light) gray regions are the domain detectable 
by SPICA for the cases where the luminosity from 
the whole pre-galactic cloud is 
$L_{\rm cl}=10^{37}\psi_{\rm III}{\rm erg~s^{-1}}$ 
($=5 \times 10^{37}\psi_{\rm III}{\rm erg~s^{-1}}$, respectively).
Pop III star-forming galaxies at redshift $z=3-10$ 
with Pop III SFR $\sim 10^{3-4} \MO{\rm yr^{-1}}$
are detectable by SPICA.
} 
\label{fig:SFRz}
\end{figure}

First, we discuss the case of star formation through the low-density 
filaments.
In this scenario, in a large pre-galactic cloud, shocked layers
formed at virialization fragment into low-density 
($\sim 10^2{\rm cm^{-3}}$) filaments, which 
again fragment at $\sim 10^7{\rm cm^{-3}}$ into star-forming cores 
of $10-40 \MO$ (Nakamura, Umemura 2002).
In the following, the mass of the cores is assumed to be $\sim 10\MO$.
Under the assumption that the relative number of cores in a specific 
evolutionary phase is proportional to the duration of that phase, 
the luminosity averaged over the lifetime of a core $\bar{L}_{\rm core}$ 
coincides with the ensemble mean luminosity of cores in a pre-galactic cloud. 
Here, the physical quantities at the fragmentation
of the low density filaments are similar to
those of the star-forming core at $n\sim 10^7{\rm cm^{-3}}$ 
in model S6 of section 2.
Then, in the following numerical evaluation, 
we use the model S6 core at $n\gsim 10^7{\rm cm^{-3}}$ 
as a model of the fragments.
The duration of the core collapse is 
$\sim \Delta t_{\rm core}=10^5 {\rm year}$ and 
the time-averaged luminosities of the strongest lines from the cores,  
which include H$_2$ rovibrational lines of  
1--0 Q(1)($2.34{\rm \mu m}$), 1--0 O(3)($2.69{\rm \mu m}$), 
1--0 O(5)($3.04{\rm \mu m}$), and H$_2$ pure rotational lines of 
0--0 S(3)($10.03{\rm \mu m}$), 0--0 S(4)($8.27{\rm \mu m}$), 
0--0 S(5)($7.07{\rm \mu m}$)), are
$\bar{L}_{\rm core} \simeq (2-6) \times 10^{33} {\rm erg~s^{-1}}$.
On the other hand, for the low-density filament 
before fragmentation into cores (model F2 in section 3), 
the strongest lines are H$_2$ pure rotational lines: 
0--0 S(0)($29.7{\rm \mu m}$), 0--0 S(1)($17.9{\rm \mu m}$), 
and HD pure rotational lines: 
0--0 R(0)($112{\rm \mu m}$), 0--0 R(1)($56.2{\rm \mu m}$)
with $\bar{L}_{\rm fil} \simeq (2-5) \times 10^{34} {\rm erg~s^{-1}}$. 
The duration of the filamentary contraction is roughly 
$\Delta t_{\rm core}=4\times 10^6 {\rm year}$.
Since the strongest lines differ between the core and the filament, 
we can judge the conditions of emission regions by the line diagnosis. 
Next, we estimate luminosity from the whole pre-galactic cloud 
using the values of luminosity from the individual cores/filaments 
given above. 
With luminosity of a line from the core $\bar{L}_{\rm core}$ and 
number of cores in cloud $N_{\rm core}$, 
the luminosity of the line from the whole cloud $L_{\rm cl, core}$ 
is described by $L_{\rm cl, core}=\bar{L}_{\rm core} N_{\rm core}$.
The total number of star formation cores $N_{\rm core}$ is 
$\psi_{III} \Delta t/M_{\rm \ast}$ using the Pop III SFR
$\psi_{III}$, 
the typical mass of formed star $M_{\rm \ast}$ and 
the lifetime of the core $\Delta t$. 
Here we assume that the star formation efficiency $\epsilon$ 
is unity because the density is sufficiently high ($n\sim 10^7{\rm cm^{-3}}$) 
for star formation.
Using the luminosity of the strongest lines 
(e.g., H$_2$ 0--0S(3) at $10.03{\rm \mu m}$ or 0--0S(5) at$7.07{\rm \mu m}$) 
from an individual core 
$\bar{L}_{\rm core}=5\times 10^{33}({\rm erg~s^{-1}})$, 
the stellar mass $M_{\rm \ast}=10\MO$, and 
the lifetime of the core $\Delta t=10^5$yr, 
the luminosity from the whole pre-galactic cloud is
\begin{equation}
L_{\rm cl, core}=\bar{L}_{\rm core}
\left(\frac{\psi_{III} \Delta t}{M_{\rm \ast}}\right) 
=5\times 10^{39} \left( \frac{\psi_{III}}{100\MO{\rm yr^{-1}}} \right) 
{\rm erg~s^{-1}}.
\end{equation}
Next we consider the contribution from filaments.
The number of filaments in a pre-galactic cloud is
$N_{\rm fil}=\psi_{III} \Delta t_{\rm fil}/\epsilon M_{\rm fil}$. 
In numerical evaluation, we use $\epsilon = 0.1$, 
which is a typical value of star formation efficiency of 
some cluster forming regions 
(e.g., Megeath et al. 1996; Kumar et al. 2004). 
Using the contraction time of filaments
$\Delta t_{\rm fil}=4\times 10^6$yr,
the mass of a single filament $M_{\rm fil}=10^5 \MO$, 
the luminosity of the strongest line (i.e., HD 0-0R(0) at $112{\rm \mu m}$) 
of a single filament 
$\bar{L}_{\rm fil}=5\times10^{34} {\rm erg~s^{-1}}$, 
the sum of luminosities from filaments is
\begin{equation}
L_{\rm cl, fil}=
\bar{L}_{\rm fil}\left(\frac{\psi_{III} \Delta t}{\epsilon M_{\rm fil}}\right)
=2 \times 10^{39} 
\left( \frac{\psi_{III}}{100\MO{\rm yr^{-1}}} \right) {\rm erg~s^{-1}}.
\end{equation}
Thus, the flux from the cloud at redshift $z$ 
\begin{equation}
F=\frac{L_{\rm cl}}{4\pi D_{1+z}^2}
\end{equation}
where $L_{\rm cl}$ is the luminosity from the whole pre-galactic cloud 
and $D_{1+z}$ is luminosity distance at redshift $z$. 
We use the cosmological parameters $H_0=70{\rm km s^{-1} Mpc^{-1}}$, 
$\Omega_{\Lambda}=0.7$ and $\Omega_{\rm M}=0.3$.
In Figure \ref{fig:SFRz} we show the redshift-SFR domain
where the H$_2$-line flux exceeds the detection limit of SPICA, 
$10^{-22}-10^{-21} {\rm W m^{-2}}$ in the mid- and far-infrared wavelengths 
with a spectroscopic resolution $R=3000$ and observation time of 10 hours 
(private communication; H. Matsuhara). 
SPICA is the most sensitive in the redshifted wavelengths of 
molecular emission for $z<10$ among next-generation facilities. 
In Figure \ref{fig:SFRz}, the top and bottom lines are drawn 
by the criterion $ F=L_{\rm cl}/4\pi D_{1+z}^2 > 10^{-22}{\rm W m^{-2}}$ 
using the luminosities from the whole cloud 
$L_{\rm cl}=10^{37}\psi_{III} {\rm erg~s^{-1}}$ and 
$=5 \times 10^{37}\psi_{III} {\rm erg~s^{-1}}$, respectively. 
With Pop III SFR as high as $\sim 10^{3}-10^{4} \MO{\rm yr^{-1}}$
at redshift $z<8-10$,
SPICA can detect the strongest H$_2$ and HD lines, such as
H$_2$: 1--0 Q(1), 1--0 O(3), 1--0 O(5), 0--0 S(0), 0--0 S(1), 
0--0 S(3), 0--0 S(4), 0--0 S(5), HD: 0--0 R(1).
Being more sensitive than SPICA for 
$\lambda<10{\rm \mu m}$, JWST can detect H$_2$ vibrational lines 
under the same condition in Figure \ref{fig:SFRz}.
However, strong H$_2$ and HD pure rotational lines are 
redshifted away from the JWST band for $z \gsim 3$.
To study the physical condition of emission regions 
using the luminosity ratio of rotational and vibrational lines, 
SPICA is more suitable. 

Next, we discuss the case of star formation 
through the high-density filaments ($ \gsim 10^6 {\rm cm^{-3}}$).
Because the formation mechanism of these high-density filaments remains 
unclear, we do not try to include the emission from the lower density region
and consider only the contribution from the high-density filaments 
(model F3 in section 3). 
The strongest emission lines in this case are very similar to those 
from spherical cores, and their time-averaged luminosities are 
$\bar{L}_{\rm fil} \gsim 10^{37} {\rm erg~s^{-1}}$ 
for filament mass of $10^5 \MO$ (see figure \ref{fig:L(time)cyl}c).  
The number of filaments in the pre-galactic cloud 
$N_{\rm fil}=(\psi_{III} \Delta t_{\rm fil}/\epsilon M_{\rm fil})$. 
Because all the matter in the filaments is dense enough to 
form stars, we assume that SFR in these filaments is unity: $\epsilon =1$. 
The sum of luminosities in a strong line over all the filaments
in the pre-galactic cloud
\begin{eqnarray}
L_{\rm cl, fil}=\bar{L}_{\rm fil}
\left(\frac{\psi_{III} \Delta t}{\epsilon M_{\rm fil}}\right)
=10^{37} \psi_{III}~{\rm erg~s^{-1}},
\end{eqnarray}
where we have used the contraction time 
$\Delta t_{\rm fil}=10^5$yr and the mass of a filament
$M_{\rm fil}=10^5\MO$ and the luminosity of the strongest lines 
(e.g., H$_2$: 1--0 Q(1), 1--0 O(3), 1--0 O(5), 0--0 S(3), 0--0 S(4), 
0--0 S(5)) 
$\bar{L}_{\rm fil}=10^{37}{\rm erg~s^{-1}}$. 
The $z-\psi_{III}$ domain for detection of the molecular emission 
by SPICA for the high-density filaments is also shown 
in Figure \ref{fig:SFRz}.   
We can see that SPICA and JWST can detect the molecular emission 
if the SFR is as high as $\sim10^{3-4}\MO{\rm yr ^{-1}}$ 
at redshift $z<7$. 

In summary, in both cases, high ($\sim 1000\MO{\rm yr^{-1}}$) Pop III SFR
at redshift $z<7$ is required for protogalaxies 
to be observable by next-generation facilities through the molecular 
emission.

How much mass is needed in these pre-galactic clouds ?
Duration of star formation in a pre-galactic cloud would be
larger than the free-fall time of the cloud,
$t_{\rm ff} \simeq 10^7 {\rm yr}$ for $\simeq 1{\rm cm^{-3}}$. 
For star formation to continue at a rate of $\psi_{III}$ 
during this period, gas mass of $\psi_{III} t_{\rm ff}$ is required.
For $\psi_{III} >10^4\MO{\rm yr^{-1}}$, 
the gas mass of a pre-galactic cloud 
exceeds $10^{11}\MO$.
Also, metallicity in the cloud must be lower than $10^{-4}-10^{-3}Z_{\odot}$ 
to cool and emit a copious amount of primordial molecular emission 
(Omukai et al. 2005).
In the standard bottom-up scenario of galaxy formation, 
when building blocks assemble into a protogalaxy as massive as 
$\sim 10^{9} \MO$,  
the volume-averaged metallicity of gas has reached 
$[{\rm Fe/H}]=-2.30$ via metal diffusion by supernovae 
(Mori et al. 2004).
Their results also show that the metal mixing proceeds
quite imhomogeneously.
Large spatial variation of metallicity ranging 
from $[{\rm Fe/H}] \approx -5$ to 0
is found in high-density ($>1 {\rm cm^{-3}}$) regions.
While the ambient gas of supernovae is polluted by metals,
most of the gas remains at low metallicity, $[{\rm Fe/H}]\sim -5$.
Within a relatively metal-polluted pre-galactic cloud,
a large number of almost metal-free regions can be contained.
If star formation is active in such very metal-poor regions,
next-generation facilities might have a chance of detecting the
H$_2$ and HD lines emitted in those regions.

\subsection{Observational Strategy for Discovering Pop III Galaxies}
Finally, we propose a strategy for discovering pre-galactic clouds that 
contain Pop III star-forming regions.
Such objects are likely to be metal-deficient.
A method for finding such objects is then
to look for the L$\alpha$-line emission,
which is expected to be emitted copiously 
($\sim 10^{42}{\rm erg~s^{-1}}$) at 
the virialization of massive ($10^{11}-10^{12}\MO$) 
pre-galactic clouds (Omukai, Kitayama 2003). 
A number of L$\alpha$ blobs and L$\alpha$ emitters have been found
in recent surveys.
Although the emission mechanism of those objects still remains disputed,
among hypotheses is the cooling radiation at the virialization
(Steidel et al. 2000, Haiman et al. 2000, Fardel et al. 2001).
Some L$\alpha$ blobs are very diffuse and in their vicinity no
bright continuum source is found.
These properties are indeed expected for the L$\alpha$-cooling objects
(Matsuda et al. 2004).

Recently, L$\alpha$ blobs that are brighter than $10^{43}
{\rm erg~s^{-1}}$ and larger than $100$kpc have been found
in and around a protocluster region at redshift $z=3.1$
(e.g., Steidel et al. 2000, Matsuda et al. 2004).
Those L$\alpha$ blobs meet the detection condition derived in section 4.2 
above, namely, being Milky-Way sized and located at $z\sim3$.
If those L$\alpha$ blobs indeed originate in the cooling radiation
of low-metallicity gas, there might be a number of Pop III star-formation
regions inside them.
As an observational strategy, we propose first
to select candidates by investigating metallicities of L$\alpha$ blobs 
with spectroscopic analysis
and then to look for the molecular lines using SPICA or JWST 
(see also Omukai, Kitayama 2003).

\section{Summary}
We have studied chemical evolution of primordial molecules 
(H$_2$, HD and LiH) in star-forming clumps 
and evaluated their line luminosities. 
We have found conditions of the star formation rate and 
the formation redshift of a Pop III star-forming 
pre-galactic cloud to be observable by SPICA and JWST.

In the case with high initial H$_2$ abundance,
HD cooling becomes effective at low density $n<10^6{\rm cm^{-3}}$
and affects the thermal evolution. 
At densities higher than the LTE critical density 
$\sim 10^5{\rm cm^{-3}}$,
temperature increases and HD is dissociated.
At high enough density ($n>10^9{\rm cm^{-3}}$),
H$_2$ dominates the cooling and HD has negligible effects 
on the evolution.
LiH cooling remains insignificant 
despite the expectation that LiH can be an important coolant 
owing to abundant LiH formation at the high density 
($n>10^{12}{\rm cm^{-3}}$; e.g., Lepp, Shull 1986; Stancil et al. 1996).
Even though the three-body LiH formation is included,
the molecular fraction of lithium remains low ($<10^{-3}$)
owing to efficient dissociation through 
${\rm LiH + H \rightarrow Li + H_2}$.

Molecular emission from a star-forming core reaches the 
maximum shortly after the formation of a central protostar.
Since the evolution has already converged by this moment, 
maximum luminosities of lines hardly depend on the initial conditions.
Most luminous lines include H$_2$ lines of 1--0 Q(1),
1--0 O(3), 0--0 S(3), and 0--0 S(4). 
Their luminosities are $\gsim 10^{35}{\rm erg~s^{-1}}$ 
for the core of $10^{3}\MO$ .
For a contracting high-density filament, 
strongest lines include H$_2$ lines of 1--0 Q(1), 1--0 O(3), 
1--0 O(5), 0--0 S(3), 0--0 S(4) and 0--0 S(5).
The luminosities of these lines are $\gsim 10^{37}{\rm erg~s^{-1}}$
for a filament of $10^{5} \MO$.
On the other hand, for a low-density filament of $10^5\MO$, 
the luminosities reach $>10^{34}{\rm erg~s^{-1}}$ for the strongest lines 
( H$_2$ lines of 0--0 S(0), 0--0 S(1), and HD lines of 0--0 R(0), 0--0 R(1) )
before fragmentation into cores. 
In protogalaxies with a high Pop III star formation rate 
$\sim 10^{3}-10^{4}\MO{\rm yr^{-1}}$ at redshift $z<7$, 
SPICA is able to detect H$_2$ lines such as 1--0 Q(1), 
1--0 O(3), 1--0 O(5), 0--0 S(0), 0--0 S(1), 0--0 S(3), 0--0 S(4),
and 0--0 S(5), and the HD line 0--0 R(1). 
Since JWST is more sensitive than SPICA in the near-infrared
wavelengths $\lsim 10 {\rm \mu m}$,
JWST also can detect the H$_2$ rovibrational lines.
SPICA, whose bands include mid- and far-infrared wavelengths,
has advantage of observing redshifted H$_2$ and HD pure 
rotational lines emitted $z>3$. 
Then, we can reveal physical conditions of star-forming regions 
by diagnosing the rotational and vibrational line ratios. 
Provided such objects exist, these facilities can directly probe  
Pop III star formation through H$_2$ lines.

The authors wish to acknowledge Andrea Ferrara, 
Daniele Galli, Takashi Hosokawa, Tohru Nagao, 
Francesco Palla, Masayuki Umemura, and Toru Yamada for valuable comments, 
and Ken-ichi Oohara and Kazuya Watanabe for continuous encouragement.
This work is supported in part by research fellowships of the Japan Society 
for the Promotion of Science for Young Scientists and 
a Grant-in-Aid for Scientific Research (16204012 KO; 14540227 RN) 
of the Ministry of Education, Culture, Science and Technology. 

\clearpage
\newpage

\newpage

\newpage

\newpage

\newpage

\newpage

\newpage

\newpage

\newpage

\newpage

\newpage

\newpage

\newpage

\clearpage
\newpage
%%%%%%%%%%%%%%%%%%%%%%%%%%%%%%%%%%%%%%%


\begin{thebibliography}{l}
\bibitem[]{}
Abel, T., Bryan, G. L., \& Norman, M. L. 2000, ApJ, 540, 39
\bibitem[]{}
Abel, T., Bryan, G. L., \& Norman, M. L. 2002, Science, 295, 93
\bibitem[]{}
Bennett,C. L., et al. 2003, ApJS, 148, 1
\bibitem[]{}
Bougleux, E., \& Galli, D. 1997, MNRAS, 288,638
\bibitem[]{}
Bromm, V., Coppi, P. S., \& Larson, R. B. 1999, ApJ, 527, L5
\bibitem[]{}
Bromm, V., Kudritzki, R. P., \& Loeb, A. 2001, ApJ, 552, 464
\bibitem[]{}
Bromm, V., Coppi, P. S., \& Larson, R. B. 2002, ApJ, 564, 23
\bibitem[]{}
Christlieb, N., et al. 2002, Nature, 419, 904
\bibitem[]{}
Ciardi, B., \& Ferrara, A. 2001, MNRAS, 324, 648
\bibitem[]{}
Ciardi, B., Ferrara, A., \& White, S. D. M. 2003, MNRAS, 344, L7
\bibitem[]{}
Fardal, M. A., Katz, N., Gardner, J. P., Hernquist, L., Weinberg, D. H., 
\& Dave, R., 2001, ApJ, 562, 605
\bibitem[]{}
Flower, D. R., Le Bourlot, J., Pineau des Forets, G., \& Roueff, E. 2000, 
MNRAS, 314, 753
\bibitem[]{}
Galli, D., \& Palla, F. 1998, A\&A, 335, 403
\bibitem[]{}
Haiman, Z., Spaans, M., \& Quataert, E. 2000, ApJ, 537, L5
\bibitem[]{}
Hutchins, J.B. 1976, ApJ, 205, 103
\bibitem[]{}
Inutsuka, S., \& Miyama, S. M. 1997, ApJ, 480, 681
\bibitem[]{}
Kamaya, H., \& Silk, J. 2002, MNRAS, 332, 251
\bibitem[]{}
Kamaya, H., \& Silk, J. 2003, MNRAS, 339, 1256
\bibitem[]{}
Kashlinsky, A., Arendt, R., Gardner, J. P., Mather, J. C., \& Moseley, S. H.
 2004, ApJ, 608, 1
\bibitem[]{}
Kumar, N. M. S., Kamath, U. S., \& Davis, C. J. 2004, MNRAS, 353, 1025
\bibitem[]{}
Larson, R. B. 1969, MNRAS, 145,271
\bibitem[]{}
Lepp, D., \& Shull, J. M. 1984, ApJ, 280, 465
\bibitem[]{}
Lin, C. C., Mestel, L., \& Shu, F. H. 1965, ApJ, 142, 1431
\bibitem[]{}
MacLow, M. M. \& Shull, J. M. 1986, ApJ, 302, 585
\bibitem[]{}
Matsuda, Y., et al. 2004, AJ, 128, 569
\bibitem[]{}
Megeath, S. T., Herter, T., Beichman, C., Gautier, N., Hester, J. J., Rayner, J., \& Shupe, D. 1996, A\&A, 307, 775 
\bibitem[]{}
Miyama, S. M., Narita, S., \& Hayashi, C. 1987a, Prog. Theor. Phys., 78, 1051
\bibitem[]{}
Miyama, S. M., Narita, S., \& Hayashi, C. 1987b, Prog. Theor. Phys., 78, 1273
\bibitem[]{}
Mizusawa, H., Nishi, R., \& Omukai, K. 2004, PASJ, 56, 487 (paper I)
\bibitem[]{}
Mori, M., Umemura, M., \& Ferrara, A. 2004, ApJ, 613, L97
\bibitem[]{}
Nagasawa, M. 1987, Prog. Theor. Phys., 77, 635 
\bibitem[]{}
Nakamura, F., \& Umemura, M. 1999, ApJ, 515, 239
\bibitem[]{}
Nakamura, F., \& Umemura, M. 2002, ApJ, 569, 549
\bibitem[]{}
Nishi, R. 2002, Prog. Theor. Phys. Supplement, 147, 85
\bibitem[]{} 
Oh, S. P., \& Haiman, Z. 2002, ApJ, 569, 558
\bibitem[]{}
Omukai, K., \& Kitayama, T. 2003, ApJ, 599, 738
\bibitem[]{}
Omukai, K., \& Nishi, R. 1998, ApJ, 508, 14
\bibitem[]{}
Omukai, K., Tsuribe, T., Schneider, R., \& Ferrara, A. 2005, 
ApJ, 626, 627
\bibitem[]{}
Palla, F., Salpeter, E. E., \& Stahler, S. W. 1983, ApJ, 271, 632
\bibitem[]{}
Penston, M. V. 1969, MNRAS, 144, 159
\bibitem[]{}
Puy, D., \& Signore, M., 1998, NewA, 3, 27
\bibitem[]{}
Ripamonti, E., Haardt, F., Ferrara, A., \& Colpi, M. 2002, MNRAS 334, 401 
\bibitem[]{}
Salvaterra, R., \& Ferrara, A. 2003, MNRAS, 339, 973
\bibitem[]{}
Sauval, A. J., \& Tatum, J. B., 1984, ApJS, 56, 193
\bibitem[]{}
Scannapieco, E., Schneider, R., \& Ferrara, A. 2003, ApJ, 589, 35
\bibitem[]{}
Schaerer, D. 2002, A\&A, 382, 28
\bibitem[]{}
Schwarzschild, M., \& Spitzer, L. 1953, Obs., 73, 77
\bibitem[]{}
Shapiro, P. R. \& Kang, H. 1987, ApJ, 318, 32
\bibitem[]{}
Songaila, A. 2001, ApJ, 561, L153
\bibitem[]{}
Stancil, P. C., Lepp, S., \& Dalgarno, A. 1996, ApJ, 458, 401
\bibitem[]{}
Steidel, C. C., Adelberger, K. L., Shapley, A. E., Pettini, M., Dickinson, M., 
\& Giavalisco, M. 2000, ApJ, 532, 170
\bibitem[]{}
Susa, H., Uehara, H., \& Nishi, R. 1996, Prog. Theor. Phys., 96, 1073
\bibitem[]{}
Susa, H., Uehara, H., Nishi, R., \& Yamada, M. 1998, Prog. Theor. Phys., 100, 
63
\bibitem[]{}
Uehara, H., \& Inutska, S. 2000, ApJ, 531, L91
\bibitem[]{}
Uehara, H., Susa, H., Nishi, R., Yamada, M., \& Nakamura, T. 1996, 
ApJ, 473, L95
\bibitem[]{}
Venkatesan, A., Tumlison, J., \& Shull, J. M. 2003, ApJ, 584, 621
\bibitem[]{}
Yamada, M. 2002, Prog. Theor. Phys. Supplement, 147, 43  
\bibitem[]{}
Yamada, M. \& Nishi, R. 1998, ApJ, 505, 148
\end{thebibliography}
\end{document}